\documentclass[%
reprint,superscriptaddress,amsmath,amssymb,aps,floats,floatfix,showpacs,prd,twocolumn,nofootinbib,nolongbibliography]{revtex4-2}
\usepackage{mathtools,needspace,enumitem,etoolbox,physics,microtype,afterpage,bigints,tabularx,xspace}
\usepackage{graphicx, dcolumn, bm, lipsum}
\usepackage[commentmarkup=footnote]{changes}
\definecolor{linkcolor}{rgb}{0.0,0.3,0.5}
\usepackage[colorlinks = true,
            linkcolor = linkcolor,
            urlcolor  = linkcolor,
            citecolor = linkcolor,
            anchorcolor = linkcolor]{hyperref}  
\usepackage{orcidlink}
\usepackage[normalem]{ulem}
\usepackage{pifont}
\usepackage{tabularx}
\usepackage{amsmath}
\usepackage{float} 
\usepackage{placeins}
\usepackage[margin=1in]{geometry}
\usepackage{array}  
\usepackage{soul}
\usepackage{aasmacros}

\usepackage{array}

\usepackage{acronym}

\begin{document}
\preprint{APS/123-QED}
\title{A Parametrized Test of General Relativity for LISA Massive Black Hole Binary Inspirals}
\newcommand{\tou}{\affiliation{Laboratoire des 2 Infinis - Toulouse (L2IT-IN2P3), Université de Toulouse, CNRS, F-31062 Toulouse Cedex 9, France}}
\newcommand{\aei}{\affiliation{Max Planck Institute for Gravitational Physics (Albert Einstein Institute), Am M\"uhlenberg 1, Potsdam 14476, Germany}}
\newcommand{\mar}{\affiliation{Department of Physics, University of Maryland, College Park, MD 20742, USA}}

\author{Manuel Piarulli~\orcidlink{0009-0009-4099-9166}}
\email{manuel.piarulli@l2it.in2p3.fr}\tou
\author{Sylvain Marsat~\orcidlink{0000-0001-9449-1071}}
\email{sylvain.marsat@l2it.in2p3.fr}\tou
\author{Elise M. S\"anger~\orcidlink{0009-0003-6642-8974}}\aei
\author{Alessandra Buonanno~\orcidlink{0000-0002-5433-1409}}\aei\mar
\author{Jan Steinhoff~\orcidlink{0000-0002-1614-0214}}\aei
\author{Nicola Tamanini~\orcidlink{0000-0001-8760-5421}}\tou

\date{\today}

\begin{abstract}

Laser Interferometer Space Antenna (LISA) observations of massive black hole binaries (MBHBs) will provide long duration inspiral signals with high signal-to-noise ratio (SNR) data, ideal for testing general relativity (GR) in the strong-field and  relativistic regime regime. 
We present an extension of the Flexible Theory-Independent (FTI) framework, adapted to gravitational waves (GWs) from MBHBs observed with LISA, to perform parametrized inspiral tests of GR.
This approach introduces generic deviations to the post-Newtonian (PN) coefficients of the frequency-domain GW phase while accounting for the time- and frequency-dependent instrument response, thus effectively identifying potential deviations from GR by constraining modifications to the PN phasing formula. Complementary analyses using Fisher matrix and full Bayesian approaches confirm that LISA observations could improve constraints on agnostic, scale-independent deviations from GR by at least two orders of magnitude compared to the most recent LIGO-Virgo-KAGRA measurements.
Since LISA’s sensitivity to different GW phases---inspiral, merger, and ringdown---varies across the MBHB parameter space with masses between $10^4$ and $10^7M_{\odot}$, the optimal regime for testing agnostic deviations is not known a priori.
Our results illustrate how the strength of these constraints depends significantly on both the total mass and the SNR, reflecting the trade-off between inspiral and merger-ringdown contributions to the observed signal.
We also investigate the interplay between inspiral-only versus inspiral-merger-ringdown analyses in constraining these inspiral deviation parameters. 
This work contributes to the development of robust tests of GR with LISA, enhancing our ability to probe the nature of gravity and BHs with GW observations.
\end{abstract}

\maketitle
\section{\label{sec:intro} Introduction}
The detection of gravitational waves (GWs) has initiated a revolutionary era for testing Einstein's theory of general relativity (GR) in the strong-field and relativistic regime regime~\cite{1995PhRvL..74.1067B, 2006CQGra..23L..37A, 2006PhRvD..74b4006A, 2014LRR....17....4W, 2016PhRvL.116v1101A, 2019PhRvL.123a1102A}.
While ground-based detectors have established significant constraints on possible deviations from GR~\cite{GWTC1, GWTC2, GWTC3}, in particular with the clearest signal to date GW250114~\cite{KAGRA:2025oiz,2025arXiv250908099T},  the forthcoming Laser Interferometer Space Antenna (LISA) mission will enable even more stringent and complementary tests, probing different source populations ~\cite{2024arXiv240207571C, 2022LRR....25....4A}.
Operating in the low-frequency band (0.1 mHz to 1 Hz), LISA will demonstrate exceptional sensitivity to signals from massive black hole binaries (MBHBs) with masses ranging from $10^4$ to $10^7 M_{\odot}$. 
These MBHBs represent particularly promising systems for testing GR due to their characteristically high signal-to-noise ratio (SNR), particularly for high-mass systems, and extended inspirals within LISA's frequency band, particularly for low-mass systems. 
Such properties enable precise measurements of the gravitational waveform's phase evolution, making them ideal candidates for detecting potential deviations from GR's predictions.

The scientific community has established various parametrized frameworks for testing GR during the inspiral phase using ground-based detectors, including the Flexible Theory-Independent (FTI) method~\cite{FTI}, the Test Infrastructure for General Relativity (TIGER)~\cite{TIGER, TIGER_prec} and the multipolar tests of GR~\cite{2018PhRvD..98l4033K, 2019PhRvD.100d4007K, 2024PhRvD.109f4036M}. 
These methodologies have traditionally prioritized phase modifications over amplitude changes, as phase deviations accumulate across orbital cycles, thereby providing particularly sensitive probes of alternative gravity theories. 
Moreover, GW detectors exhibit less sensitivity to the signal's amplitude compared to its phase evolution~\cite{GWTC3}.

Another extensively investigated approach is the parameterized post-Einsteinian (ppE) framework, which accommodates generic deviations in both the phase and amplitude~\cite{ppe, 2016PhRvD..94h4002Y, 2017PhRvD..96h4039C, 2021PhRvD.103d4024P} of GW signals. 
The FTI, TIGER, multipolar test and ppE frameworks focus on the inspiral part of the signal. 
Recently, other approaches based on the Effective-One-Body waveform models have considered parametrized deviations in the merger-ringdown as well~\cite{Maggio:2022hre, Toubiana:2023cwr, Pompili:2025cdc}.

In this paper, we present a comprehensive framework for conducting parametrized inspiral phasing tests of GR using LISA observations of MBHBs.
Specifically, our methodology builds upon and extends the FTI framework, adapting it to the space-based detector context and to the unique properties of LISA sources.
This allows us to account for LISA's distinctive characteristics, including its time-dependent antenna response.
Through complementary Fisher matrix studies and full Bayesian analyses, we demonstrate that LISA will substantially improve current constraints on agnostic deviations from GR, with a projected enhancement of at least two orders of magnitude compared to the most recent LVK analysis of the GWTC\text3.0 catalog~\cite{GWTC3}.

It is important to emphasize that LISA and ground-based detectors probe gravitational dynamics at different curvature scales. For theories where the strength of the deviation from GR depends on the curvature, the high-mass systems observed by LISA will see a suppression of the effect, and we expect the hierarchy to be reversed, with ground instruments performing better.
In this paper, all direct comparisons between the two experiments should be understood as referring to agnostic, scale-independent deviations from GR.

Furthermore, we conduct a systematic investigation of how signal morphology across the parameter space influences our capacity to constrain GR deviations. 
Our analysis encompasses low-mass systems ($\sim 10^4 M_{\odot}$), where the inspiral phase dominates the signal in LISA's band, to high-mass binaries ($\sim 10^7 M_{\odot}$), where the merger portion predominantly contributes to the observed signal. 
This comprehensive coverage enables a deeper understanding of how the different evolutionary phases of binary systems contribute to non-GR constraints.
Since the class of deviations from GR considered is limited to the inspiral for practical reasons, rather than being justified by theory, it is important to understand where the information comes from.
While one might naively expect the inspiral part to dominate, it is the high-SNR merger-ringdown part that, while not modified away from GR, help to constrain masses and spins, thus tightening the bounds.

The paper is organized as follows:
in Sec.~\ref{sec:parametrized_test} we introduce the parametrized inspiral test of GR suitable for LISA MBHBs observations; in Sec.~\ref{sec:method} we briefly present the setup for our analyses; 
in Sec.~\ref{sec:results} we apply the GR-test method to synthetic GW signals of MBHBs. 
This allows us to assess the robustness of the presented method. 
Finally, in Sec.~\ref{sec:conclusions} we summarize our findings and discuss future improvements. 
Technical details on the post-Newtonian (PN) stationary phase approximation (SPA) are provided in Appendix~\ref{app:appendix_spa}.
Appendix ~\ref{app:appendix_alpha} examines the impact of parameters that control the decay of GR inspiral corrections to zero as the binary approaches merger.

\section{Parametrized inspiral test of general relativity \label{sec:parametrized_test}}

In this work, we will focus on binary black hole systems with no orbital eccentricity. In GR, gravitational signals from quasi-circular BHBs depend on a set of intrinsic parameters 
$\boldsymbol{\lambda} = \{m_1, m_2, \boldsymbol{S}_1, \boldsymbol{S}_2\}$, with $m_i$, $\boldsymbol{S}_i$ the compact objects masses and spins; and extrinsic parameters 
$\boldsymbol{\xi} = \{\iota, \varphi, \alpha, \delta, \psi, d_L, t_c\}$, where $\{\iota, \varphi\}$ are the polar angles of the line of sight towards the observer in a source frame defined from the orbit of the binary at a reference time, $\{\alpha, \delta\}$ the sky location in detector frame, $\psi$ is the polarization angle, $d_L$ the luminosity distance, and $t_c$ the time of arrival at the detector.
For non-precessing spins (aligned/antialigned with orbital angular momentum $\boldsymbol{L}$), the relevant spin component is $\chi_i = \boldsymbol{S}_i \cdot \boldsymbol{L}/(\boldsymbol{L}|m_i^2)$ in units where $G=c=1$ and the intrinsic parameters simplify to $\boldsymbol{\lambda} = \{m_1, m_2, \chi_1, \chi_2\}$.
We introduce additional parameters  for convenience: the mass ratio $q = m_1/m_2 \geq 1$, the symmetric mass ratio $\eta = q/(1+q)^2$, the total mass of the system $M = m_1 + m_2$, the chirp mass $\mathcal{M} = \eta^{3/5}M$, the PN spin parameter~\footnote{This PN spin parameter is the combination that appears at the leading 1.5PN order in the PN SPA phasing formula~\cite{Poisson:1995ef}, which motivates its use for inspiral-dominated signals.} $\chi_{\text{PN}} = (\eta/113)[(113q + 75)\chi_1 + (113/q + 75)\chi_2]$, and $\chi_{-} = (m_1\chi_1 - m_2\chi_2)/M$.

In this work, we closely follow the FTI pipeline, originally developed for LVK observations and adapted here to the LISA context; we refer the reader to Ref.~\cite{FTI} for further details.

\subsection{Post-Newtonian inspiral phase deviations}

The parametrized test is performed in the inspiral phase of the signal, when the compact objects are sufficiently separated, and the  PN formalism accurately describes compact binaries' dynamics~\cite{2024LRR....27....4B, 1995PhRvL..74.3515B, 1995PhRvD..52..821K, 2001PhLB..513..147D, 2005PhRvD..71l4004B, 2006PhRvD..74j4034B, 2009PhRvD..79j4023A, 2013CQGra..30m5009B, 2014PhRvD..89f4058D, 2015CQGra..32s5010B, 2023PhRvL.131l1402B}. 
The PN formalism provides an analytic approximation where metric components and gravitational fields are perturbatively expanded in powers of the velocity parameter $v/c$ (and equivalently, the parameter $(G M/(r c^2))^{1/2}$), with terms scaling as $(v/c)^n$ referred to as ``order $(n/2)$PN''.

With these assumptions, the orbital phase of the system is then analytically derivable from the flux-balance equation in PN theory, and the frequency-domain gravitational waveform is deduced from the SPA~\cite{1999PhRvD..59l4016D, 2000PhRvD..61j4015T, 2009PhRvD..80h4043B}. 

In the following, we restrict ourselves to non-precessing spins and circularized binaries and decompose the GW in spin-weighted spherical harmonics as
\begin{equation}
    h_+ - i h_\times = \sum_{\ell \geq 2} \sum_{m=-\ell}^{+\ell} {}_{-2}Y_{\ell m} h_{\ell m} \,,
\end{equation}
and we consider the Fourier-domain modes $\tilde{h}_{\ell m}(f) = A_{\ell m}(f) \exp[i \psi_{\ell m}(f)]$.

For a given $(l,m)$-mode, the structure of the GR phase in the PN expansion, using the SPA with an analytic integration of the phase and of the time-to-frequency correspondence, is expressed as~\cite{2009PhRvD..80h4043B, 2024LRR....27....4B}:
\begin{align}
    \psi^{\mathrm{GR}}_{lm}(f) &= \frac{3}{128\eta v^5}\frac{-m}{2}
    \left[ \sum_{n=0}^{7} \psi^{\mathrm{PN}}_n v^n + \sum_{n=5}^{6} \psi^{\mathrm{PN}}_{n(\ell)} v^n \log v \right] \nonumber\\
    &\quad + [\text{const}] + [\text{linear term}],
\label{eq:psi_GR}
\end{align}
where $v = (-2\pi f M / m)^{1/3} = (M \omega)^{1/3}$ is the PN expansion parameter corresponding to the orbital frequency $\omega$ (recall that $m<0$), and $\psi^{\mathrm{PN}}_n$, $\psi^{\mathrm{PN}}_{n(\ell)}$ are $(n/2)$-PN coefficients dependent on the binary's intrinsic parameters, including logarithmic terms denoted by $\ell$.
Explicit expressions up to 3.5PN order are available in Ref.~\cite{2024LRR....27....4B}.
Higher-order PN phasing terms beyond 3.5PN exist for comparable-mass binaries but are not yet implemented in parametrized tests of GR and are therefore not included here. 
For a comprehensive review of PN theory for GWs, we refer the reader to Ref.~\cite{2014LRR....17....2B}.

In the above, we left out a constant term and a linear-in-$f$ term, corresponding to the alignment degrees of freedom, comprising a phase shift and time shift; they are given explicitly in Appendix~\ref{app:appendix_spa} (see Eq.~\eqref{eq:psi22_align_explicit}).
In modified theories of gravity, the PN coefficients may differ from those predicted by GR~\cite{2016PhRvD..94h4002Y, 2018GReGr..50...46B, 2018PhRvD..98h4042T}. 
In particular, the balance equation driving the orbital phase evolution can be modified both because of a change in the binary's orbital energy and in the energy flux far from the system.

To probe potential deviations during the inspiral phase, we introduce an additional term to the frequency-domain phase of the waveform of the form:
\begin{equation}
    \begin{split}
        \delta\psi_{lm}(f) = & \frac{3}{128\eta v^5}\frac{-m}{2}\left[\sum_{n=-2}^{7}\delta\psi_n v^n + \right. \delta\psi_{4,\kappa_s}v^4 \\ 
        & + \delta \psi_{6,\kappa_s}v^6 \left. + \sum_{n=5}^{6}\delta\psi_{n(\ell)} v^n \log{v} \right],
    \end{split}
    \label{eq:dPSI}
\end{equation}
where different mode phases are all affected through the same quantity, with a simple $m/2$ prefactor and a rescaling of the relation between $v$ and the Fourier frequency $f$. 
We ignore again a constant and linear term corresponding to a possible change in alignment.

The inclusion of the negative PN order $n = - 2$ is motivated by certain alternative theories of gravity: for instance, theories involving scalar fields predict dipolar radiation, leading to deviations at $n = - 2$, where $\delta\psi_{lm} \propto v^{-7}$, which becomes particularly significant at low frequencies.
Additionally, even more negative PN orders are expected to result from astrophysical or environmental effects. 
An $n = - 8$ term may be associated with line-of-sight acceleration due to third-body interactions~\cite{2011PhRvD..83d4030Y, 2017PhRvD..95d4029B}, accretion~\cite{2014PhRvD..89j4059B, 2020ApJ...892...90C}, or a variation of the gravitational constant~\cite{2010PhRvD..81f4018Y, 2023PhRvD.107f4073B}.
Furthermore, the impact of dynamical friction on the system could arise at $n = -5.5$ \cite{2014PhRvD..89j4059B, 2021PhRvL.126j1105T}.
The detailed impact of such effects on LISA MBHBs will be investigated in future work. 
As, previously mentioned, GW detectors are more sensitive to phase evolution compared to the signal amplitude, we therefore also ignore amplitude modifications in this work~\cite{2019PhRvD.100j4001T}.

It is also useful to single out a parameter $\kappa$ characterizing the spin-induced quadrupolar deformation of a compact object. 
For Kerr BHs, this parameter takes the canonical value $\kappa = 1$~\cite{Hansen:1974zz,PhysRevLett.26.331, 2015PhRvL.114o1102G}, consistent with the no-hair theorem, which uniquely determines all multipole moments in terms of the mass and spin. 
In contrast, for other compact objects such as neutron stars (NS) or more exotic compact objects (ECOs), the value of $\kappa$ depends on the internal structure and composition, and can differ significantly from unity~\cite{2012arXiv1211.6299P, 2012PhRvL.108w1104P, 2018CQGra..35n5010H, PhysRevD.55.6081, 2014PhRvL.112v1101H, 2019PhRvD..99d4001B, 2020JCAP...11..033C, 2016CQGra..33b5005U}.

In a binary system, we define parameters $\kappa_1$ and $\kappa_2$ for the individual components, and introduce the symmetric combination $\kappa_s = (\kappa_1 + \kappa_2)/2$, which enters the waveform through spin-squared effects at the 2nd and 3rd PN orders, specifically in the phase contribution $\psi^{\mathrm{PN}}_{4-6, \kappa_s}$~\cite{2009PhRvD..79j4023A, 2015CQGra..32s5010B}. 
Higher-order contributions from $\kappa_s$ and $\kappa_a$ also exist, but are not included in the present analysis.

Measuring deviations of $\kappa_s$ from unity provides a direct probe of the nature of the compact objects involved, offering a potential test of the BH hypothesis and a way to constrain ECOs~\cite{2017PhRvL.119i1101K, 1998PhRvD..57.5287P, 1999ApJ...512..282L, PhysRevD.52.5707}.
While NS are expected to exhibit values of $\kappa$ as large as $\sim 10$--$14$~\cite{2012arXiv1211.6299P, 2012PhRvL.108w1104P, 2018CQGra..35n5010H}, they are not relevant for this study, as we focus on binary systems in a mass range incompatible with the presence of NS.

Following the conventions of Ref.~\cite{FTI}, we define each deviation parameter as a fractional correction to its corresponding PN coefficient evaluated in GR:
\begin{subequations}
\begin{align}
    \delta\psi_n & \equiv \delta\hat{\varphi}_n \psi^{\mathrm{PN}}_n, 
    \label{eq:dpsi_n} \\
    \delta\psi_{n(\ell)} & \equiv \delta\hat{\varphi}_{n(\ell)} \psi^{\mathrm{PN}}_{n(\ell)},
    \label{eq:dpsi_logn} \\
    \delta\psi_{n,{\kappa_s}} & \equiv \delta\kappa_s \psi^{\mathrm{PN}}_{n, \kappa_s}
\end{align} \label{eq:def_deltaphi} 
\end{subequations}
For PN orders where the GR coefficients vanish (e.g., $n = -2, 1$), the parameter $\delta \hat{\varphi}_n$ represents an \emph{absolute} deviation rather than a fractional one. 
In current LVK analyses~\cite{GWTC3}, deviations are tested only in the non-spinning contributions to the PN coefficients $\psi^{\mathrm{PN}}_n$, while the spin-dependent terms are held fixed at their GR values. 
This restriction is motivated by strong parameter degeneracies and enables direct comparison with complementary tests, such as the TIGER framework~\cite{TIGER}. 
The FTI method~\cite{FTI}, as initially proposed, formally incorporates the full spin-dependent expressions for $\psi^{\mathrm{PN}}_n$; its more recent implementation allows for the same set-up as in LVK analyses.
In this work, we follow the original convention of FTI and consider deviations in the full spin-dependent expressions of $\psi^{\mathrm{PN}}_n$ rather than restricting modifications to only the non-spinning part.

Consequently, Eq.~\eqref{eq:dPSI} can be rewritten as:
\begin{align}
        \delta\psi_{lm}(f) = & \frac{3}{128\eta v^5}\frac{m}{2} \bigg[ \sum_{\substack{n=0 \\ n \neq 1}}^{7} \delta\hat{\varphi}_n \psi^{\mathrm{PN}}_n v^n + \delta\kappa_s (\psi^{\mathrm{PN}}_{4,\kappa_s}v^4 \nonumber\\ 
        & + \psi^{\mathrm{PN}}_{6,\kappa_s}v^6) + \sum_{n=5}^{6} \delta\hat{\varphi}_{n(\ell)} \psi^{\mathrm{PN}}_{n(\ell)} v^n \log{v} \nonumber\\
        & + \sum_{\substack{n=-2,1 }} \delta\hat{\varphi}_n v^n \bigg].
    \label{eq:dPSI_phi}
\end{align}

Deviation coefficients can be treated as either free parameters or can be mapped to specific predictions from alternative theories that admit PN-like perturbative expansions~\cite{FTI}.

Alternative theories of gravity introduce deviations in multiple PN coefficients away from their GR values~\cite{2016PhRvD..94h4002Y, 2018GReGr..50...46B, 2018PhRvD..98h4042T}.
However, since these coefficients are typically correlated in their joint posterior distribution, allowing multiple deviations to vary simultaneously leads to significantly weaker constraints. 
To mitigate this degeneracy, and in line with standard practice in LVK analyses~\cite{GWTC3}, we adopt here a single-parameter approach in which one deviation coefficient is varied at a time while all others are fixed to their GR values.
Although single-parameter tests may not pinpoint specific theories, they remain sensitive to beyond-GR effects~\cite{2013PhRvD..87j2001S, 2018PhRvD..97d4033M, 2021PhRvD.104b4060P, 2022PhRvD.105l4047P}.
Marginally tighter constraints in the multi-parameter regime can be obtained by reparametrizing correlated deviations into orthogonal combinations using Principal Component Analysis (PCA)~\cite{2013CQGra..30b5011P, 2022PhRvD.105h4062S, 2023GReGr..55...55S, 2024PhRvD.109d4036D, 2024ResPh..5707407N, Mahapatra:2025cwk, 2025arXiv250908099T}.

\subsection{Implementation details}

To extend this formalism to post-inspiral portions of the waveform (merger and ringdown), we must ensure continuity and consistency. 
Following Ref.~\cite{FTI}, we require that the non-GR parametrization satisfies:

\begin{enumerate}
    \item The phase of the early-inspiral (low-frequency) waveform is given by:
    \[
    \psi_{l m}(f) = \psi_{l m}^{(\text{GR})}(f) + \delta \psi_{l m}(f)
    \]

    \item The phase of the post-inspiral (high-frequency) waveform is given by the GR one:
    \[
    \psi_{l m}(f) = \psi_{m}^{(\text{GR})}(f).
    \]

    \item The two regimes are connected smoothly, in a sense to be made precise below.
\end{enumerate}

In order to limit the deviations exclusively to the inspiral phase and to enforce a smooth transition back to GR, we employ a Planck tapering function $ W(f; f^\text{tape}, \Delta f^\text{tape}) $ (see Eq. (7) in Ref.~\cite{2010CQGra..27h4020M}). 
The tapering frequency $f^{\text{tape}}$ is chosen as a fraction of the $(2,2)$-mode peak frequency $f^{\text{peak}}_{22}$, typically $f^{\text{tape}}_{22} = \alpha f^{\text{peak}}_{22}$, with $\alpha = 0.35$ \cite{FTI}. 
The tapering interval is chosen to be $[f^{\text{start-tape}}, f^{\text{end-tape}}]$ with $f^{\text{start-tape}} = 0.9 f_{22}^{\text{tape}}$ and $f^{\text{end-tape}} = 1.1 f_{22}^{\text{tape}}$.
Consistent with Ref.~\cite{FTI}, the tapering function will not be applied to the phase itself but to its second derivative with respect to frequency $\psi_{\ell m}''(f)$, which is related to the chirp rate of the inspiral (see Appendix~\ref{app:appendix_spa}).
This choice is made in order to taper a quantity most closely related to the modified physics of the system, instead of the phase itself, which has extra phase and time alignment degrees of freedom (corresponding to integration constants when going from $\psi_{\ell m}''$ to $\psi_{\ell m}$).

To obtain the total phase correction for the (2,2)-mode, we divide the frequency grid into three regions:
\begin{itemize}    
    \item For $f \in [f^{\text{start-tape}}, f^{\text{end-tape}}]$, we construct the phase correction $\delta\psi_{22}(f)$ by integrating the second-order frequency derivative $\delta\psi''_{22}(f)$ multiplied by the tapering window $W(f)$, as follows:
\begin{equation}
    \delta\psi_{22}(f) = \int_{f_{\text{int}}}^f \! df' \int_{f_{\text{int}}}^{f'} \! df'' \, 
    \delta\psi''_{22}(f'') W(f'').
    \label{eq:phase_correction}
\end{equation}
Here, the reference frequency for the integration $f_{\text{int}} \equiv f^{\text{end-tape}}$ is chosen to ensure that $\delta\psi_{22}(f_{\text{int}}) = 0$, $\delta\psi_{22}'(f_{\text{int}}) = 0$. This choice guarantees that the phase correction and its derivative go to zero at the end of the window, for a smooth attachment to the post-inspiral part of the waveform.
    \item For $f \in [f^{\text{min}}, f^{\text{start-tape}})$, $\delta \psi_{22}(f)$ is computed as in Eq.~(\ref{eq:dPSI_phi}), where $f^{\text{min}} = 10^{-5} \text{Hz}$, with an additional constant and linear term ensuring $C^1$-continuity at $f^{\text{start-tape}}$.
    \item For $f \in (f^{\text{end-tape}}, f^{\text{max}}]$, $\delta \psi_{l m}(f) = 0$, where $f^{\text{max}} = 0.5 \text{Hz}$.
\end{itemize}
One further needs to specify a reference frequency to align the waveform, that is to say a frequency where time and phase values are imposed, $\psi_{22}(f_{\rm ref}) = \psi_{\rm ref}$, $\psi_{22}'(f_{\rm ref}) = 2\pi t_{\rm ref}$. 
We use the peak frequency of the (2,2)-mode (although our implementation allows for a generic choice); since it lies in the range where $\delta \psi$, $\delta \psi'$ are zero by construction, the modified waveform is already aligned. 
In practice, this means that the modified waveforms remain aligned (coincident in time and phase) with the GR signals at the peak of the waveform.

The phase modification to different harmonics is then directly obtained via a rescaling:
\begin{equation}\label{eq:freq_scaling}
    \delta \psi_{\ell m} \left( \frac{m f}{2} \right) = \frac{m}{2} \delta \psi_{22} (f) \,,
\end{equation}
reflecting the fact that in the PN regime, all mode phases represent multiples of the orbital phase. 
In particular, what we call the tapering frequency $f^{\text{tape}}$ corresponds actually to an orbital frequency $\omega^{\text{tape}} = \pi f^{\text{tape}}$ at which the corrections are smoothly tapered off, which in turn corresponds to different Fourier frequencies for each modes $f_{lm}^{\text{tape}} = m/2 f^{\text{tape}}$.

The choice of $\alpha$ and the tapering width $\Delta f^{\text{tape}}$ is purely phenomenological. 
Adopting the simple choice $\alpha = 0.35$ ensures that our analysis remains conservative, terminating inspiral deviations well in advance of reaching the merger.
In future work, we may revisit this choice, particularly in light of recent works~\cite{Elise2024}, where GR deviations are permitted to extend up to $f_{22}^{\text{peak}}$, corresponding to $\alpha = 1$. 
We will illustrate the impact of varying $\alpha$ on our analysis in Appendix~\ref{app:appendix_alpha}. 
In brief, we find that low- and intermediate-mass systems are insensitive to the choice of $\alpha$, while high-mass systems exhibit a stronger dependence, with larger values of $\alpha$ generally leading to tighter constraints on high-PN order deviations.

\begin{figure*}[ht!]
    \centering
    \includegraphics[width=\textwidth]{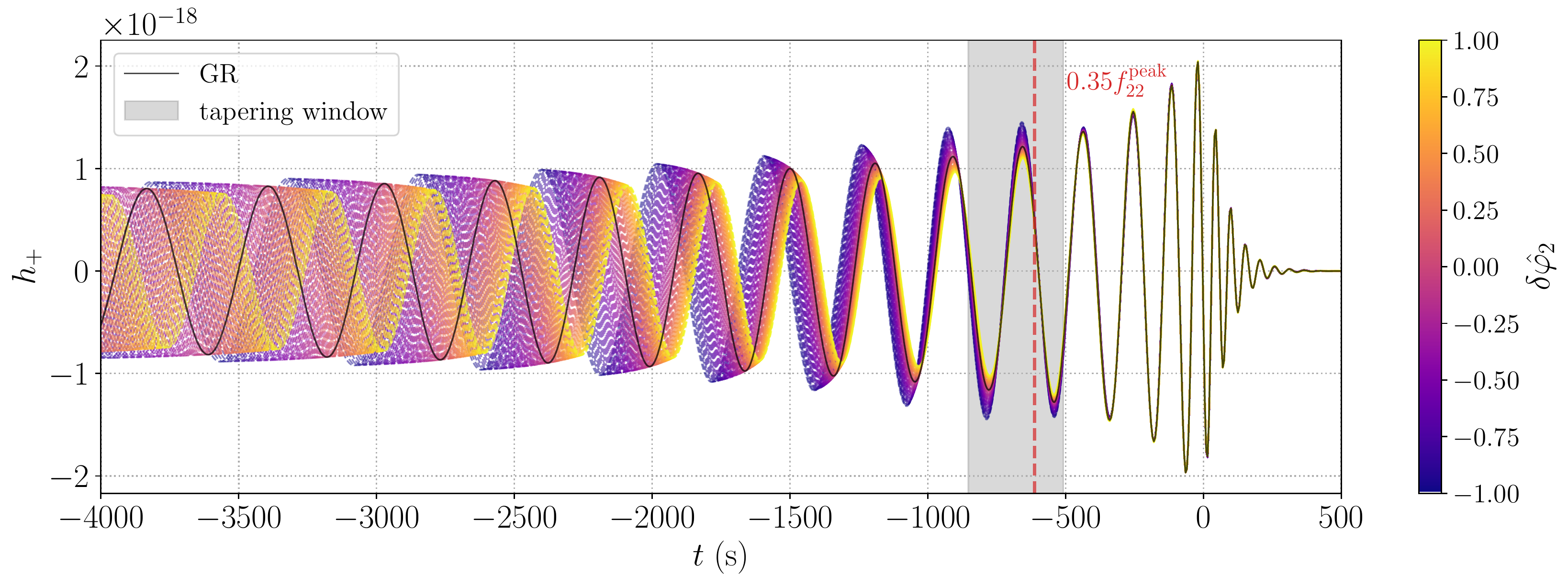}
    \caption{An example of the plus polarization, $h_+$, of the dominant $(2,2)$-mode GR waveform is shown in black, together with its non-GR counterparts in color for several values of $\delta\hat{\varphi}_2 \in [-1,1]$. The vertical red dashed line marks the time corresponding to the tapering frequency, $f^{\text{tape}}_{22} = 0.35 f^{\text{peak}}_{22}$, around which the non-GR deviations are gradually switched off. The shaded grey region indicates the window over which this smooth tapering is applied. By construction, the GR and non-GR waveforms coincide in the post-tapering region.} \label{fig:wf_time}
    \end{figure*}

Compared to the uniform frequency grid used in Ref.~\cite{FTI} for LVK signals, our approach partitions the frequency grid into three regions. 
This is motivated by the nature of MBHBs signals, which are persistent signals in LISA frequency bands and accumulate many waveform cycles, making full-grid integration potentially computationally expensive.
In our implementation, only the intermediate tapering region $f \in [f^{\text{start-tape}}, f^{\text{end-tape}}]$ requires a numerical integration (as imposed by the presence of the taper $W$, even though each term in $\delta\psi_{22}''$ is a power law in $v$ or $f$). 
This is done by building a cubic interpolating spline of the $W(f'')\delta\psi_{22}''(f'')$ data, before integrating twice to build the integrated function as a piecewise 5-th order polynomial. 
The number of points used for integration is $\sim 10^3$. 
This ensures a numerical accuracy well below the phase accuracy floor of $10^{-4} \mathrm{rad}$, which the code \texttt{lisabeta}~\cite{lisabeta} employs as its default target for cubic spline phase interpolation errors in the sparse representation of the GR waveform.
This integration step could in principle employ an even sparser representation, but the added cost of our implementation with respect to GR signals remains at the millisecond level, independent of the system. 
For reference, the computational cost of a GR waveform generation, typically ranges from a few hundred microseconds up to about one millisecond, depending on the signal duration. 
Generating the non-GR signal introduces an overhead of approximately one millisecond—a $50$--$100\%$ relative increase that remains computationally negligible due to the low baseline cost.

In Fig.~\ref{fig:wf_time}, we show the plus polarization, component of a (2,2)-mode only MBHB signal, for the parameterized non-GR waveform and its GR counterpart. 
Of particular interest is the behavior around the tapering frequency $f_{22}^{\text{tape}}$, chosen as $0.35f_{22}^{\text{peak}}$. 
To explore potential deviations from GR, we plot waveforms with a set of different values in the deformation parameter $\delta\hat{\varphi}_2$, allowing it to range from $-1$ to $1$.
This systematic parameter variation illustrates how GR deviations are tapered around $f_{22}^{\text{tape}}$. 
The non-GR waveform exhibits significant dephasing from GR during the early inspiral phase ($t < t^{\text{tape}}$), but these modifications are suppressed as the frequency approaches $f_{22}^{\text{tape}}$, ensuring a smooth convergence to the GR prediction in the late inspiral regime. 
As previously mentioned, the two waveforms are aligned at the peak frequency of the (2,2)-mode.
This figure also shows that the modification introduced is not only a change in phase, as it also translates in a different time-frequency track, resulting in an apparent stretch of the waveform.

\section{\label{sec:method} Analysis Setup}

\subsection{Parameter estimation for LISA MBHBs}

The presented framework is implemented in the \texttt{lisabeta} package~\cite{lisabeta}, suitable for comprehensive LISA responses while accounting for both LISA's motion and time delay interferometry (TDI)~\cite{2021LRR....24....1T}.
We employ first-generation ``noise-orthogonal'' TDI variables $A$, $E$, and $T$, under the approximation of equal and static LISA arm lengths~\cite{1999ApJ...527..814A, 2021LRR....24....1T, Hartwig:2022yqw}.
The implementation of the instrumental response of LISA in the Fourier domain is done using the same approach as in~\cite{,lisabeta}; the only modification being that the phase modifications $\delta \psi_{\ell m}$ lead to changes in the time-of-frequency relation required to evaluate the instrument's geometry $\delta t_{\ell m}(f) = 1/(2\pi ) \times d \delta \psi_{\ell m}/df$, which is computed by taking a derivative of a cubic interpolating spline.
This gives us the TDI signals in a sparse Fourier-domain amplitude/phase representation as
\begin{equation}
    \tilde{s}_k = \sum_{\ell m} \mathcal{T}_k^{\ell m}(f) \tilde{h}_{\ell m}(f) \exp \left[ i\delta \psi_{\ell m}(f)\right] \,.
\end{equation}
The $\mathcal{T}_k^{\ell m}$ are TDI transfer functions for individual modes, written as function of frequency here with the time-dependent quantities evaluated at the modified time-of-frequency $t_{\ell m}(f) + \delta t_{\ell m}(f)$. We refer to~\cite{lisabeta} for further notations and definitions.

In this work, we apply our parametrized tests to MBHB systems using two different waveform models, which allows us to test the robustness of our results to the choice of model used. 
For our main results, we use the aligned-spin inspiral-merger-ringdown phenomenological model \texttt{IMRPhenomXHM}~\cite{XHM_wf, 2020PhRvD.102f4002G};
for comparison, we also perform selected analyses with the aligned-spin effective-one-body (EOB) waveform model \texttt{SEOBNRv5HM\_ROM}~\cite{2017PhRvD..95d4028B, 2020PhRvD.101l4040C}.
The two waveform models include different sets of spherical harmonic modes: $(l, m) = (2, 2), (2, 1), (3, 3), (3, 2), (4, 4)$ for \texttt{IMRPhenomXHM} and $(l, m) = (2, 2), (2, 1), (3, 3), (3, 2), (4, 4), (4, 3) (5, 5)$ for \texttt{SEOBNRv5HM\_ROM}.

Waveform models are only approximations to the true astrophysical signals, an issue of potentially crucial importance for the high-SNR LISA MBHB signals.
For the purpose of this study, we assume perfect agreement between the model and the physical waveform, and all our analyses use the same model to inject a simulated signal and to recover its parameters, thereby neglecting systematics that may arise from waveform inaccuracies~\cite{Toubiana:2023cwr}, as well as from neglected environmental effects~\cite{Garg:2024qxq}.
We leave for future work the impact of waveform systematics on the test of GR explored in this work.

While recent developments include a TIGER framework based on the IMRPhenomX waveform family that also incorporates precession~\cite{TIGER_prec}, in this study we restrict our analysis to spin-aligned and quasi-circular systems for simplicity. 
The waveform models and parameter-estimation infrastructure we use are currently implemented and validated only for aligned-spin, quasi-circular binaries. 
Extending the analysis to include precessing or eccentric systems would require additional code development, waveform integration, and validation, which lies beyond the scope of the present work.
The inclusion of more astrophysically complete waveform models—accounting for precession, eccentricity, and other effects—is left for future work as well.

The single-source signal-to-noise ratio (SNR), $\rho$, is expressed as:
\begin{equation}
    \rho^2 = \sum_{k} (s_k|s_k)_k,
    \label{eq:snr_tdi}
\end{equation}
where $k$ runs over the noise-orthogonal TDI channels $k=(A, E, T)$ and $s_k$ is the signal in the corresponding channel. 
For any two time-dependent functions $a(t)$ and $b(t)$, their inner product is defined as:
\begin{equation}
    (a|b)_k = 4\mathrm{Re} \int_{0}^{\infty} \mathrm{d}f \, \frac{\widetilde{a}^*(f) \widetilde{b}(f)}{S_{n,k}(f)},
    \label{eq:inner_prod}
\end{equation}
where $S_{n,k}(f)$ denotes the power spectral density (PSD) for each TDI channel $k$.

Under the assumptions of stationary, Gaussian noise independent across TDI channels and characterized by PSDs $S_{n,k}(f)$, the likelihood of observing a set of data streams $(d_k)_k$ given a signal model $s_k(\bm{\theta})$ is expressed as
\begin{equation}
    \ln \mathcal{L} = -\frac{1}{2} \sum_{k=A,E,T} \left( s_k(\bm{\theta}) - d_k \,\middle|\, s_k(\bm{\theta}) - d_k \right)_k,
    \label{eq:loglike}
\end{equation}
where $\bm{\theta}$ denotes the set of physical parameters defining the waveform, and the inner product is defined according to Eq.~\eqref{eq:inner_prod}. 
In the above, the data set $(d_k)_k$ consist of a true signal $s_k(\bm{\theta}_0)$, corresponding to the actual source parameters $\bm{\theta}_0$, superimposed with a noise realization $n_k$ in each TDI channel:
\begin{equation}
    d_k = s_k(\bm{\theta}_0) + n_k.
\end{equation}.

We will first compute parameter uncertainties using the Fisher information matrix methodology.
Under our assumption of additive Gaussian stationary noise and independent channels, the Fisher information matrix takes the form:
\begin{equation}
F_{ij} = \sum_{k=A,E,T} (\partial_i s_k|\partial_j s_k)_k \,,
\end{equation}
where $\partial_i$ denotes the partial derivative with respect to the $i$-th component of the parameter vector $\boldsymbol{\theta}$. 
Within this framework, the likelihood function can be approximated as:
\begin{equation}\label{eq:lnL_fisher}
\ln \mathcal{L} \approx -\frac{1}{2} F_{ij}\Delta\theta^i\Delta\theta^j
\end{equation}
where $\Delta\theta^i$ represents the deviation from the true signal. 
The parameter uncertainties are estimated by computing the Gaussian covariance matrix $\boldsymbol{\Sigma} = \mathbf{F}^{-1}$ from the inverse Fisher matrix. 
In our analyses, we ensured numerical stability of the Fisher matrix inversion by checking the criterion $\max_{i,j} \left| I_{\text{num}}^{ij} - \delta^{ij} \right| < 10^{-4}$,
following~\cite{2005PhRvD..71h4025B, 2011PhRvD..83d4036S, 2020PhRvD.102h4056M}. 
$I_{\text{num}}^{ij}$ denotes the matrix obtained from the product of the Fisher matrix and its inverse, which should be the identity in the ideal case, and $\delta_{ij}$ is the Kronecker delta.

In practice, we found that it is not uncommon for the resulting approximated Gaussian distributions to extend beyond the allowed physical range for the parameters, most notably with BH spins outside the range $[-1, 1]$. 
In order to eliminate this issue, we draw $\sim 10^4$ samples from a multivariate Gaussian distribution corresponding to Eq.~\eqref{eq:lnL_fisher}, and we enforce physical plausibility by rejecting samples for which the spins fall outside the range $[-1, 1]$. 
The $90\%$ quantile constraints are then derived from the distribution of the accepted samples. 
This samples-based procedure is computationally light and improves the robustness of Fisher analyses, which by construction do not incorporate prior bounds.

To complement and validate our Fisher analysis, we also conduct full Bayesian inference studies on a subset of events, using the \texttt{lisabeta} package \cite{lisabeta}. 

Bayesian inference allows us to update our knowledge about the parameters $\bm{\theta}$ through the posterior distribution
\begin{equation}
    p(\bm{\theta}|d) = \frac{p(d|\bm{\theta})\,p(\bm{\theta})}{p(d)},
\end{equation}
where $p(\bm{\theta})$ represents prior beliefs about the parameter values, and $p(d)$ is the model evidence or marginal likelihood. 
As our aim is to characterize parameter estimation rather than perform model comparison, we treat $p(d)$ as a normalization factor and do not compute it explicitly. 

To simplify and accelerate the likelihood computation, we adopt the \textit{zero-noise approximation}, in which we set the noise contributions $(n_k)_k$ to zero. 
This approach is sufficient for understanding parameter degeneracies and likelihood shapes, and already offers a marked improvement over Fisher matrix-based estimates. 
Extensions to include realistic noise realizations are reserved for future investigations.

In practice, for both Fisher matrix and likelihood evaluations, the frequency-domain waveforms and inner products are evaluated numerically on a discrete frequency grid. 
To improve computational efficiency while retaining sensitivity to the most informative parts of the signal, we adopt a non-uniform frequency sampling scheme: we use fine sampling at low frequencies, where the signal's long inspiral is rich in phase evolution, and coarse sampling at high frequencies, where the signal's short merger carries less phase cycles. 
This adaptive strategy allows for accurate likelihood evaluations with reduced computational cost, and is especially useful for long-duration signals like those observed by LISA.

To perform the posterior sampling, we employ the \texttt{ptemcee} package~\cite{2016MNRAS.455.1919V}, which implements a parallel-tempered Markov Chain Monte Carlo algorithm based on an ensemble sampler~\cite{2010CAMCS...5...65G, 2013PASP..125..306F}. 
This approach enhances exploration of complex, multi-modal parameter spaces by running multiple chains at different ``temperatures" and periodically exchanging information between them. In addition, \texttt{lisabeta} implements additional jump proposals aiming at resolving multimodal structure in the sky~\cite{lisabeta}, although not the focus of the present study.
We then obtain one-dimensional posteriors for the deviation parameters by marginalizing the 12-D posterior over all other parameters. 
As mentioned in Sec.~\ref{sec:parametrized_test}, we allow only one deviation parameter at a time, analyzing each independently. 
This results in eleven separate runs per system.

\begin{figure}[t!]
    \centering
    \includegraphics[width=\columnwidth]{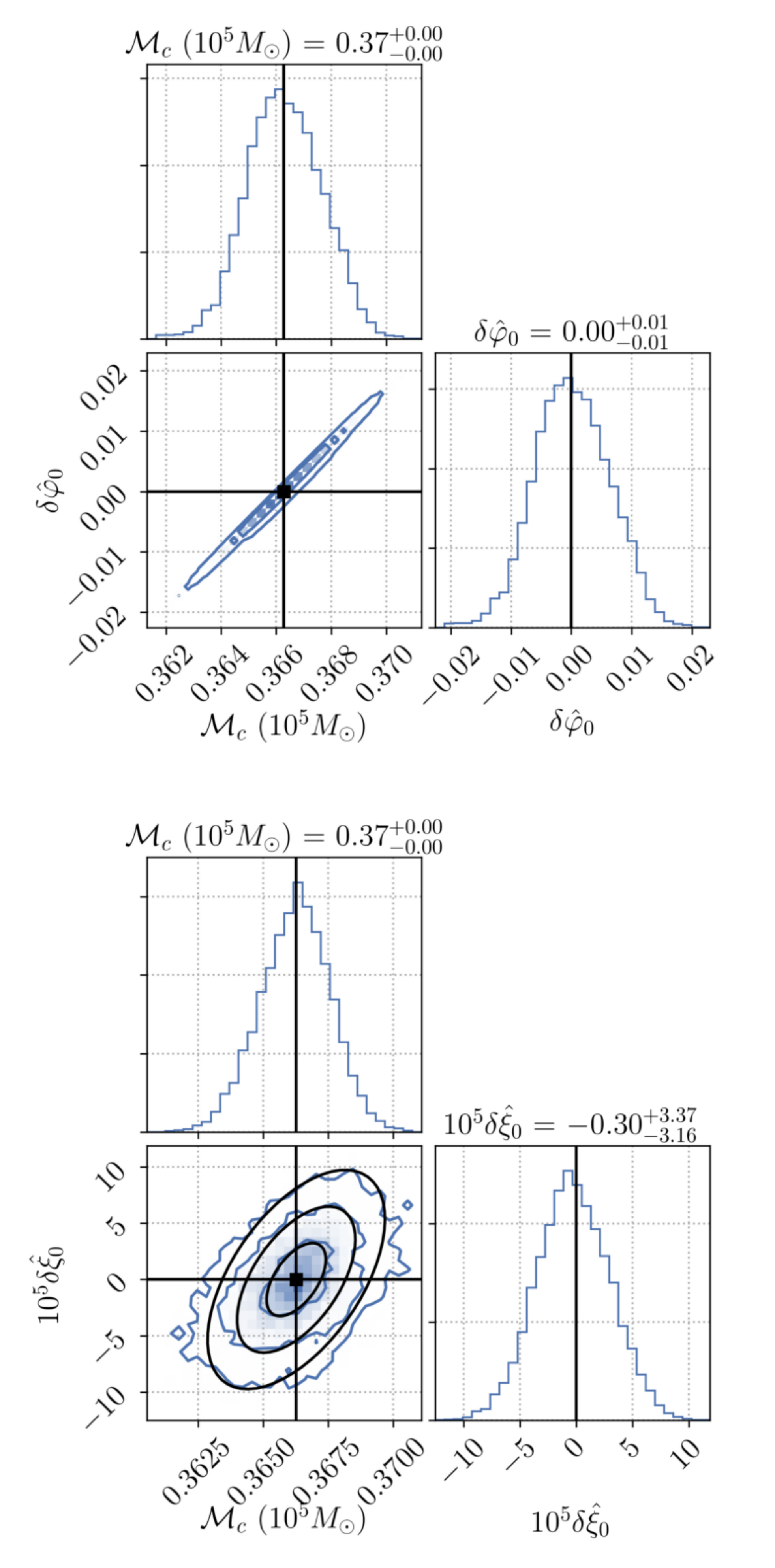}
    \caption{The top panel shows the two-dimensional posterior distribution (blue) for the chirp mass $\mathcal{M}_c$ and the non-GR parameter $\delta\hat{\varphi}_0$ for an example system. Owing to the strong degeneracy between these parameters, as described by Eq.~\eqref{eq:Mc_chi0_deg}, the Fisher-predicted contours (black) are not visible—reflecting a significant underestimation of the uncertainties and the failure of the Fisher approximation to capture the parameter correlations. In contrast, the bottom panel shows the posterior distribution for $\mathcal{M}_c$ and the reparametrized deviation parameter $\delta\hat{\xi_0}$, where the Fisher contours are in good agreement with the full Bayesian posterior, demonstrating the improved performance of the Fisher approach under this reparametrization.} \label{fig:corner_Mc_fti_dxi0v_Mc_fti_dchi0v}
\end{figure}

\subsection{Reparametrization of the 0PN-deviation}
\label{sub:chi0_Mc}

At low frequencies, the phase of the GR waveform is dominated by the 0PN term, driven in GR by the chirp mass alone.
In our effective modified-gravity waveform model, the phase contribution from the 0PN term is given by
\begin{equation}
\begin{split}
\Psi^{\text{0PN}}_{\ell m}(f) 
&= \frac{3}{128 \eta v^5} \frac{m}{2} (1 + \delta \hat{\varphi}_0) \psi_0^{\text{GR}} \\
&= C_0 \frac{1 + \delta \hat{\varphi}_0}{\mathcal{M}_c^{5/3}} f^{-5/3} , 
\end{split}
\end{equation}

where $\psi_0^{\text{GR}}=1$, and $C_0$ is a constant that does not depend on the intrinsic parameters of the binary.
As already shown in Ref.~\cite{Elise2024} (see Sec. IVC therein, and also  Sec. IVC in Ref.~\cite{FTI}), this expression shows that the deviation parameter $\delta \hat{\varphi}_0$ is degenerate with the chirp mass of the binary. 
As a result, these two quantities are strongly correlated, particularly for low-mass systems that are dominated by the early inspiral part of the signal. 
Assuming that the true signal is consistent with GR and that the true chirp mass is $\mathcal{M}_c^{\text{true}}$, this correlation takes the form
\begin{equation}
\delta \hat{\varphi}_0 = \left( \frac{\mathcal{M}_c}{\mathcal{M}_c^{\text{true}}} \right)^{5/3} - 1.
\label{eq:Mc_chi0_deg}
\end{equation}
    
Due to this strong degeneracy, we found that for inspiral dominated signals ($M = 10^4, 10^5 M_{\odot}$), a naive application of the Fisher matrix formalism leads to significant inaccuracies in the estimated uncertainties on $\delta \hat{\varphi}_0$. 
In particular, comparison with full Bayesian inference shows that the Fisher-estimated posterior width for $\delta \hat{\varphi}_0$ is underestimated by approximately an order of magnitude. 
This behaviour did not arise for high mass systems ($M = 10^6, 10^7 M_{\odot}$), where the higher SNR and the presence of the merger in the signals helps in breaking this degeneracy.
As illustrated in the top panel of Fig.~\ref{fig:corner_Mc_fti_dxi0v_Mc_fti_dchi0v}, the black contours—representing the Fisher estimate—are not visible at all due to their severe underestimation.
This discrepancy indicates that the Fisher approximation does not adequately capture the parameter correlations induced by the chirp-mass degeneracy at the 0PN order. 
To mitigate this issue, we reparametrize the deviation parameter as

\begin{equation}
\delta \hat{\xi_0} \equiv \delta \hat{\varphi}_0 + \left( \frac{\mathcal{M}_c}{\mathcal{M}_c^{\text{true}}} \right)^{5/3} - 1,
\end{equation}
\begin{figure*}[t!]
    \centering
    \includegraphics[width=\textwidth]{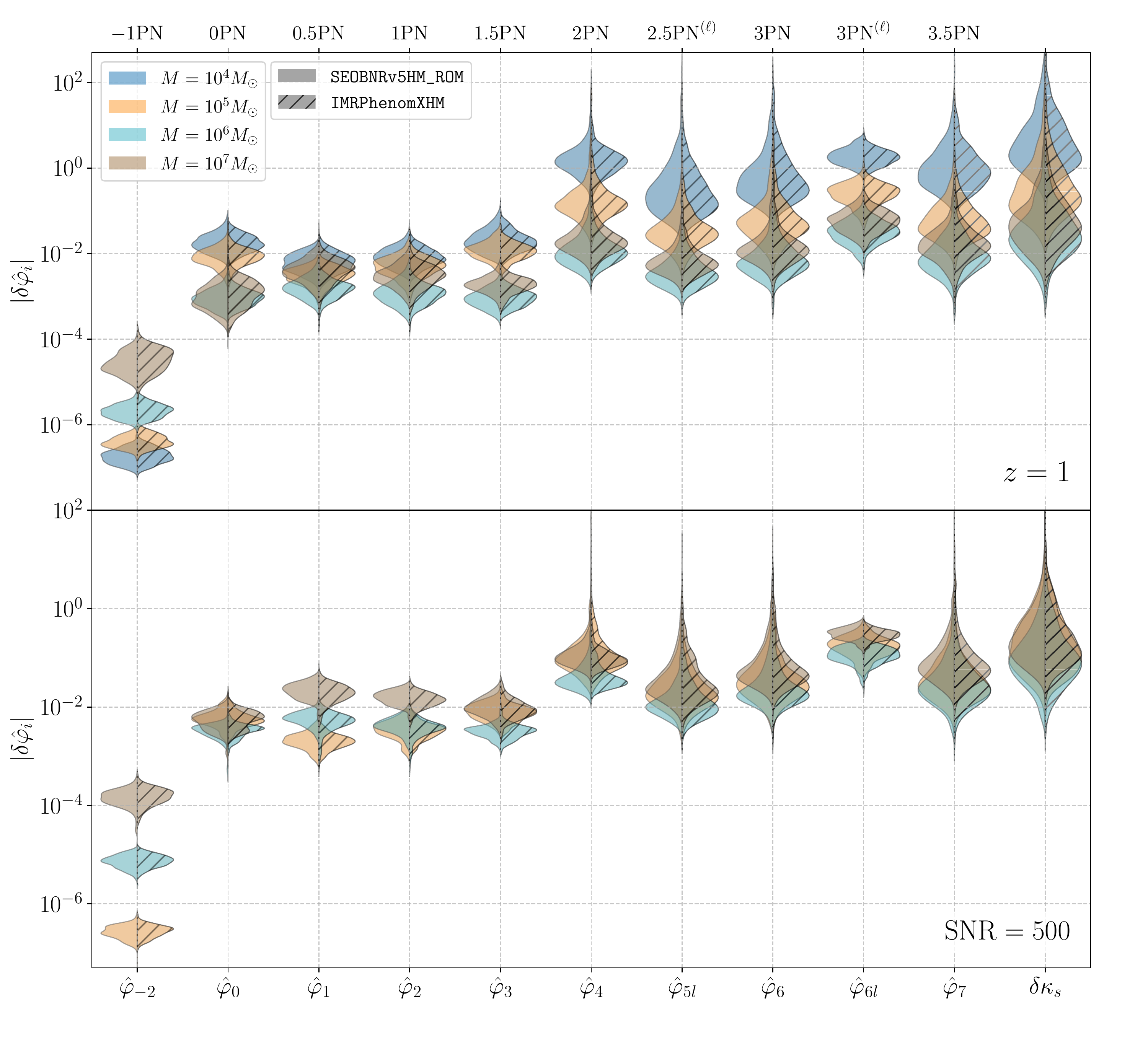}
    \caption{\textbf{Top panel:} Distribution of $90\%$ upper bounds for non-GR deviation parameters $|\delta\hat{\varphi}_i|$ across different total mass systems ($M = 10^4, 10^5, 10^6, 10^7\,M_\odot$) at redshift $z=1$.
    \textbf{Bottom panel:} Distribution of $90\%$ upper bounds for non-GR deviation parameters $|\delta\hat{\varphi}_i|$ parameters across mass systems ($M = 10^5, 10^6, 10^7\,M_\odot$) with redshift adjusted to maintain fixed SNR $= 500$.
    Both analyses use 500 Fisher matrix computations with randomly sampled spins and mass ratio. Left , undashed (right, dashed) side of each distribution uses \texttt{SEOBNRv5HM\_ROM} (\texttt{IMRPhenomXHM}) waveform model for the analysis.} \label{fig:violin_diffM_comparison}
\end{figure*}
which isolates the part of $\delta \hat{\varphi}_0$ that is independent of the leading-order correlation with the chirp mass. 
Using this reparametrization, we compute a Fisher matrix for the set of parameters $\bm{\theta}' = (\delta \hat{\xi_0}, \bm{\theta}_{\rm GR})$ instead of the original set $\bm{\theta} = (\delta \hat{\varphi}_0, \bm{\theta}_{\rm GR})$. The uncertainty $\sigma(\delta \hat{\varphi}_0)$ is then obtained by sampling from a multivariate Gaussian distribution constructed from the Fisher covariance matrix for the parameter set $\bm{\theta}'$ with the transformed variable $\delta \hat{\xi_0}$, mapping back the samples to the original parameter $\delta \hat{\varphi}_0$, and computing the standard deviation from the $\delta \hat{\varphi}_0$ samples. 
This approach is computationally cheap and yields uncertainty estimates in better agreement with the full Bayesian posteriors, as it more accurately captures the effect of parameter correlations within the Fisher framework, as shown in the bottom panel of Fig.~\ref{fig:corner_Mc_fti_dxi0v_Mc_fti_dchi0v}.

When sampling in the transformed parameter space, the reparameterization to $\delta \hat{\xi_0}$ preserves the original prior distributions due to the unit Jacobian determinant of the transformation.
This allows existing prior specifications on $\delta \hat{\varphi}_0$ (uniform in our case) to be applied directly to $\delta \hat{\xi_0}$ and vice-versa, without reweighting of the samples.

Such a remapping can also, in principle, help the convergence of the sampler used for the Bayesian inference. In our case, we found that sampling in $\delta \hat{\varphi}_0$ or sampling in $\delta \hat{\xi_0}$ and mapping samples back to the $\delta \hat{\varphi}_0$ variable were giving consistent results. This indicates that the Bayesian sampler achieved robust convergence even in the degenerate case, at least in the examples considered.

\section{\label{sec:results} Results}

\subsection{Fisher analyses}
\label{sec:results_fisher}

Using the methodology outlined in Sec.~\ref{sec:method}, we now present the results of our Fisher matrix analysis for MBHB systems with total masses of $\{10^4, 10^5, 10^6, 10^7\} M_\odot$ at fixed redshift $z=1$.
For each total mass, we perform 500 independent Fisher computations, randomizing over intrinsic and extrinsic parameters to explore a broad region of the MBHB parameter space.
Specifically, component spins ($\chi_1$, $\chi_2$) are drawn uniformly from the interval $[-1, 1]$, while the mass ratio ($q$) is logarithmically sampled within $[1, 8]$.
The angle pairs $(\iota, \phi)$ and $(\lambda, \beta)$ are sampled to be uniformly distributed over the sphere, while the polarization angle $\psi$ is drawn uniformly from the interval $[0, \pi]$.

This comprehensive analysis is carried out using both the \texttt{IMRPhenomXHM} and \texttt{SEOBNRv5HM\_ROM} waveform models to assess model dependence and robustness.

In the top panel of Fig.~\ref{fig:violin_diffM_comparison}, we show the distributions of $90\%$ upper bounds on the absolute value of the deviation parameters $|\delta\hat{\varphi}_i|$ across various PN orders. 
We find a good consistency between the two waveform families, which reinforces the reliability of the constraints.

The projected uncertainties span a wide dynamic range, from $10^{-7}$ up to a few hundred (recall that the quantities $\delta \hat{\varphi}_i$ are defined as  relative deviations from the GR value, see Eq.~\eqref{eq:def_deltaphi}), highlighting the interplay between total mass, signal morphology, and SNR in determining parameter precision.
At positive PN orders, lower-mass systems ($10^4$–$10^5 M_\odot$), while having long-duration inspirals with many cycles in band, exhibit relatively large uncertainties due to their intrinsically lower SNR. 
Conversely, higher-mass binaries benefit from higher SNR signals and from capturing both inspiral and merger-ringdown phases within LISA's most sensitive band; this helps estimating parameters such as masses and spins, which in turn refines the constraint on the deviation from GR, that would otherwise be affected by correlations with those parameters. 
Systems around $10^6 M_\odot$ seem to represent an optimal regime, achieving the tightest constraints due to this favorable balance between signal strength and information content.

For the negative PN order $n=-2$, this trend reverses. 
The $\delta\hat{\varphi}_{-2}$ term, being proportional to $v^{-7}$, dominates the early inspiral and is best constrained by low-mass, inspiral-dominated sources with long in-band evolution.

Figure~\ref{fig:violin_diffM_comparison} lower panel provides a complementary analysis where the SNR is fixed at 500 for the masses $[10^5, 10^6, 10^7] M_\odot$ by appropriately adjusting each system’s redshift. 
Note that we exclude $10^4 M_\odot$ systems in this fixed-SNR analysis, as achieving an SNR of 500 is not expected for these systems (see,  e.g., Ref.~\cite{2024arXiv240207571C}).
With such a fixed SNR, the difference between inspiral- and merger-dominated systems is now caused by the difference in signal morphology, shedding a new light on the interplay between mass and SNR. 
Inspiral-dominated binaries, such as those with $10^5 M_\odot$, now achieve better precision across low-PN orders ($n < 3$) than their higher-mass counterparts. 
This confirms that the degradation in parameter estimation for higher-mass binaries stems from the limited information available in the low-frequency, inspiral phase. 
When moving to high-frequencies contribution $ n \geq 3$ in the PN-phasing formula, the merger-dominated higher-mass signals ($10^6$--$10^7 M_\odot$) show comparable or even better constraints.
We interpret this as a combination of the importance of information on masses and spins from the merger, together with the fact that these corrections are most important in the late inspiral shortly before merger.

Overall, our results demonstrate that MBHBs observed with LISA offer significant improvements in testing GR. 
The projected bounds on deviations improve by at least two orders of magnitude over current ground-based observations. Notably, constraints on $\delta\hat{\varphi}_{-2}$ improve by approximately four orders of magnitude for systems in the $[10^4, 10^5] M_\odot$ mass range, when compared to the latest LVK results from GWTC-3.0~\cite{GWTC3} for BBH systems, and by two orders of magnitude when comparing to constraints from the binary neutron star merger GW170817~\cite{GWTC2,2019PhRvL.123a1102A}. 
So far, the most stringent constraints are obtained from binary pulsars in the large-separation regime where current bounds reach the level of $\delta\hat{\varphi}_{-2} < 2.6 \times 10^{-10}$~\cite{2014arXiv1402.5594W, 2010PhRvD..82h2002Y, 2020PhRvD.101j4011N, PhysRevX.11.041050}
These constraints are, however, source-dependent. 
As emphasized in~\cite{2016PhRvL.116x1104B}, bounds from pulsars and BHBs are complementary, since NS may not undergo scalarization in certain beyond-GR theories (e.g., shift-symmetric Einstein-scalar-Gauss-Bonnet gravity~\cite{2015PhRvL.115u1105B, 2016CQGra..33e4001Y, 2016PhRvD..93b4010Y}), while BH can carry scalar modifications. 

MBHBs observed by LISA are therefore expected to provide precise constraints that complement those derived from binary pulsar observations. 
Previous Fisher matrix studies of LISA sources within the ppE framework have investigated dipolar radiation and other PN corrections~\cite{2017PhRvD..96h4039C, 2021PhRvD.103d4024P}. Our results for individual sources are broadly consistent with those of Ref.~\cite{2017PhRvD..96h4039C}. 

However, the analysis in~\cite{2021PhRvD.103d4024P} employed full MBHB population models, which typically yield tighter projected constraints due to the combination of information from a large number of observations.
In contrast, we focus on a representative set of individual benchmark sources to assess the fundamental sensitivity limits, deferring a comprehensive population-based analysis to future work.

Our projected constraints on the deformation in the spin-induced quadrupole parameter $\delta\kappa_s$ from individual MBHB sources span $90\%$ upper bounds between $\sim 10^{-3}$ and several hundred, with the tightest constraints arising from high-mass systems ($M \sim 10^6 - 10^7 M_{\odot}$). 
These represent a substantial improvement over current ground-based GW constraints: the latest LVK analysis reported bounds from combined events of $\delta\kappa_s < 6.66$ (positive values) and $\delta\kappa_s= -16.0^{+13.6}_{-16.7}$ (symmetric bounds) at $90\%$ credibility~\cite{GWTC3}. 
Our projections on $\delta\kappa_s$ yield tighter constraints than those of Ref.~\cite{2020CQGra..37t5019K} due to the inclusion of the merger-ringdown in our analyses. 
When imposing a maximum frequency before the merger phase, however, the results become comparable (see Sec.~\ref{sec:cut_f}).
The enhanced sensitivity of LISA to massive compact object binaries thus offers the prospect of stringent constraints on ECOs, potentially distinguishing between BH and alternative compact objects with great confidence.

To contextualize our results, we compare them with projected constraints from other current and future GW detectors. 
For ground-based observatories, Einstein Telescope (ET)~\cite{2008arXiv0810.0604H} is projected to improve constraints on deviations in the inspiral coefficients by approximately two orders of magnitude over Advanced LIGO's first observing run, as demonstrated through analysis of a GW150914-like signal~\cite{2025arXiv250312263A}. 
Our LISA projections yield comparable or superior constraints: slightly more stringent for all positive PN orders, and approximately two orders of magnitude stronger for the $-1$ PN order, owing to LISA's sensitivity to the extended inspiral phase of MBHB signals, particularly at lower masses $10^4 - 10^5 M_\odot$.

Similar conclusions were reached by Ref.~\cite{2020PhRvL.125t1101G}, which analyzed a GW150914-like signal in the ppE framework across multiple detector generations. 
They found that LISA alone provides competitive constraints on negative PN-order parameters, while Cosmic Explorer (CE)~\cite{2019BAAS...51g..35R} dominates at positive PN orders. 
Crucially, multiband observations combining LISA with CE improve bounds by more than an order of magnitude across several PN orders, demonstrating the complementary nature of space- and ground-based detectors.

For the $-1$ PN order specifically, which parameterizes dipolar radiation, the landscape of projected constraints varies significantly across detector types and source populations. 
LISA's access to the early inspiral regime, dominated by low-frequency content, enables particularly strong constraints on stellar-mass black hole binaries, with projected uncertainties of $\sigma(\hat{\varphi}_{-2}) \sim 10^{-8}$. 
ET is able to achieve even tighter constraints of $\sigma(\hat{\varphi}_{-2}) \sim 2 \times 10^{-9}$ for optimal neutron star binary signals~\cite{2021PhRvD.104h4008Z}. 
The most stringent projected constraints come from multiband observations combining a far-future deci-Hz detector such as DECIGO~\cite{2021PTEP.2021eA105K} with CE, reaching $\sigma(\hat{\varphi}_{-2}) \sim 3 \times 10^{-13}$~\cite{2020MNRAS.496..182L}.

The constraints reported in this work correspond to the analysis of a single MBHB signal. 
In the presence of multiple detections of MBHBs, as expected over LISA's mission duration, these constraints could be improved by combining measurements hierarchically.
However, the degree of improvement critically depends on the population characteristics and on the assumptions made in the hierarchical analyses. 
In the ideal case of $N$ independent sources with identical deviations for all systems, statistical uncertainties are expected to scale down as $1/\sqrt{N}$, potentially leading to significant improvements in the bounds on beyond-GR parameters. 
With a more conservative approach, considering deviations drawn from a common distribution, collective constraints might be dominated by the few best events with limited additional benefit from the population.
We leave inference over realistic populations for future studies.

\subsection{\label{sec:cut_f} Impact of the merger-ringdown on inspiral constraints}

\begin{figure*}[t!]
    \centering
    \includegraphics[width=\textwidth]{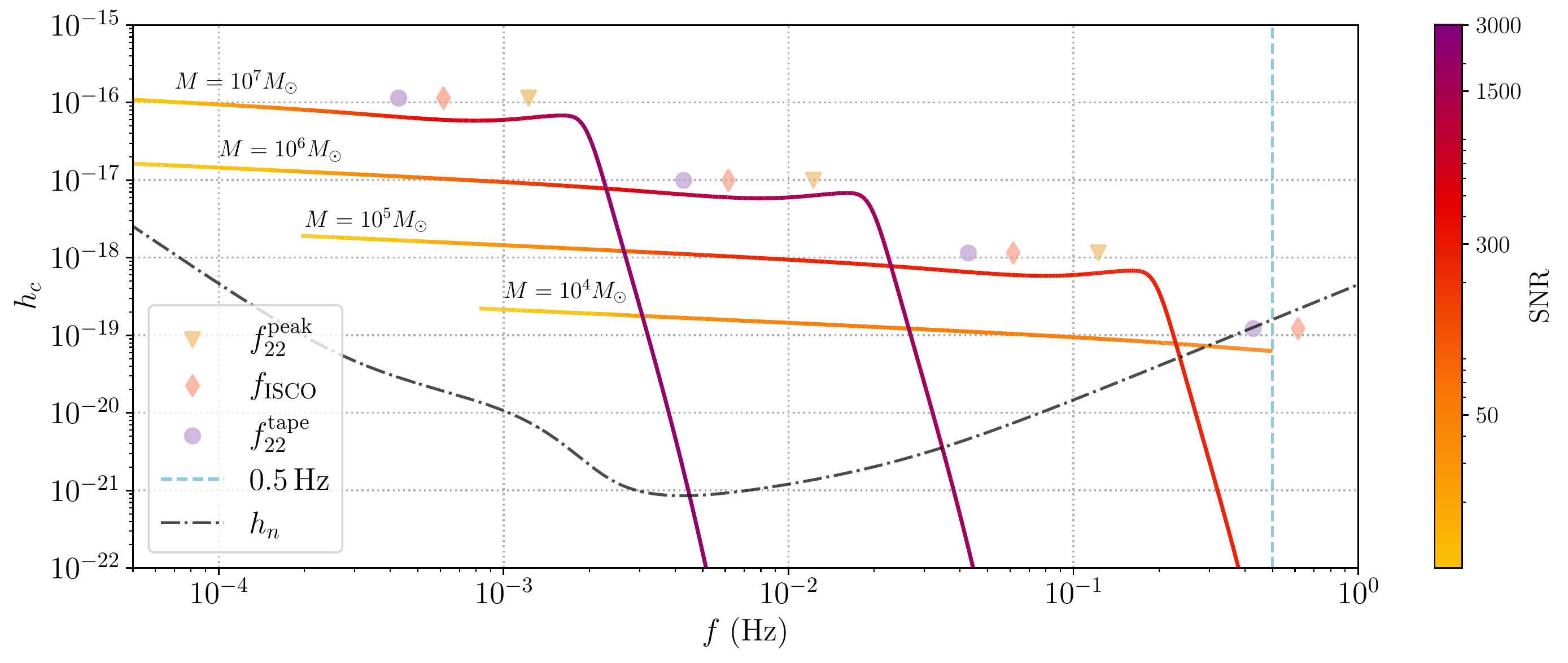}
    \caption{Characteric strain, $h_c$, of GW signals from MBHBs as a function of frequency. The systems analyzed all have $q=3$, $z=1$, and $\chi_{1,2} = 0.5, 0.2$, and are shown with their total detector-frame masses, $M$, indicated on top of each colored trace. The trace colors correspond to the accumulated SNR at each frequency. The dot-dashed black line represents LISA’s instrumental characteristic noise, $h_n$, the dashed blue vertical line denotes the nominal maximum frequency for LISA, $f = 0.5$~Hz, and the yellow triangle (salmon diamond, violet circle) represent the frequencies at (2,2)-peak (ISCO, (2,2)-tape) per each system.} \label{fig:mbh_amp_cutf}
    \end{figure*}
    
To investigate which portion of the GW signal contributes most to the parameter uncertainties discussed in the previous section, we perform a series of Fisher analyses with varying upper frequency cutoffs, using the \texttt{IMRPhenomXHM} waveform model.

Specifically, we vary the maximum frequency, $f^\text{{max}}$, used in the analysis. 
In practice, this corresponds to modifying the definition of the inner product in Eq.~\eqref{eq:inner_prod} as follows:
\begin{equation}
    (a|b)_k = 4\mathrm{Re} \int_{0}^{f^{\text{max}}} \mathrm{d}f \, \frac{\widetilde{a}^*(f) \widetilde{b}(f)}{S_{n,k}(f)},
    \label{eq:inner_prod_f_max}
\end{equation}
which directly affects the likelihood evaluation in Eq.~\eqref{eq:loglike}.
Given the independence of Fourier frequency bins under the stationarity assumption, this corresponds to marginalizing the likelihood over data with frequencies $f>f^{\text{max}}$, which is a statistically well-defined operation. 
Note that this cut in the data itself differs from changing the tapering frequency $f_{\text{tape}}^{22}$ for the deviation from GR; we investigate the effect of the latter choice in App.~\ref{app:appendix_alpha}.

The analysis employs four distinct frequency cutoffs, each providing insights into different phases of the gravitational wave signal. 
The 0.5\,Hz cutoff captures the full waveform, including the complete merger-ringdown phase, up to the point where the instrumental noise PSD is rising and cutting off contributions to the likelihood. 
The cutoff at $ f_{22}^{\text{peak}} $, corresponding to the peak of the dominant $(2,2)$-mode, extends up to the merger, capturing both the late inspiral and merger dynamics. 
The more conservative $ f_{22}^{\text{tape}} = 0.35f_{22}^{\text{peak}}$ cutoff, for some systems falls below the innermost stable circular orbit (ISCO) frequency~\cite{2024PhRvD.110b4013K}, $ f_{\text{ISCO}}$, and is used to isolate the inspiral regime.

It is important to note that this cut is applied directly on the data frequency content, see Eq.~\eqref{eq:inner_prod_f_max}, and should not be confused with a cut in the orbital frequency, or at some chosen time pre-merger.
Such a cut would instead follow a different scaling across modes and depend on the source's parameters: for each multipolar mode $(\ell, m)$ in the waveform, the frequency $f^{\text{max}}$ corresponds to an orbital frequency scaled with $m/2$ according to Eq.~\eqref{eq:freq_scaling}.
In particular, the merger of the $(2,1)$ harmonic might still be included for a cut at $ f_{22}^{\text{peak}} $; while for higher-$m$ harmonics, the cut at $f^{\text{max}}$ corresponds to a lower orbital frequency and earlier time than for the $(2,2)$-mode, excluding a longer portion of the signal. 
Our choice was made to remain statistically consistent and simplifies the analysis, but is conservative  by potentially excluding valid contributions from higher modes at frequencies beyond the cutting frequency.

Figure~\ref{fig:mbh_amp_cutf} illustrates the inspiral-merger-ringdown (IMR) tracks of MBHBs as a function of frequency. 
The plot also highlights the different cutoffs points chosen for each system, showcasing the morphology of signals across different mass ranges. 
The MBHBs systems picked for this illustration have total masses of $\{10^4, 10^5, 10^6, 10^7\} M_\odot$, redshift $z = 1$, component spins $\chi_{1,2} = 0.5, 0.2$, and mass ratio $q = 3$.

\begin{figure*}[htbp]
    \includegraphics[width=\textwidth]{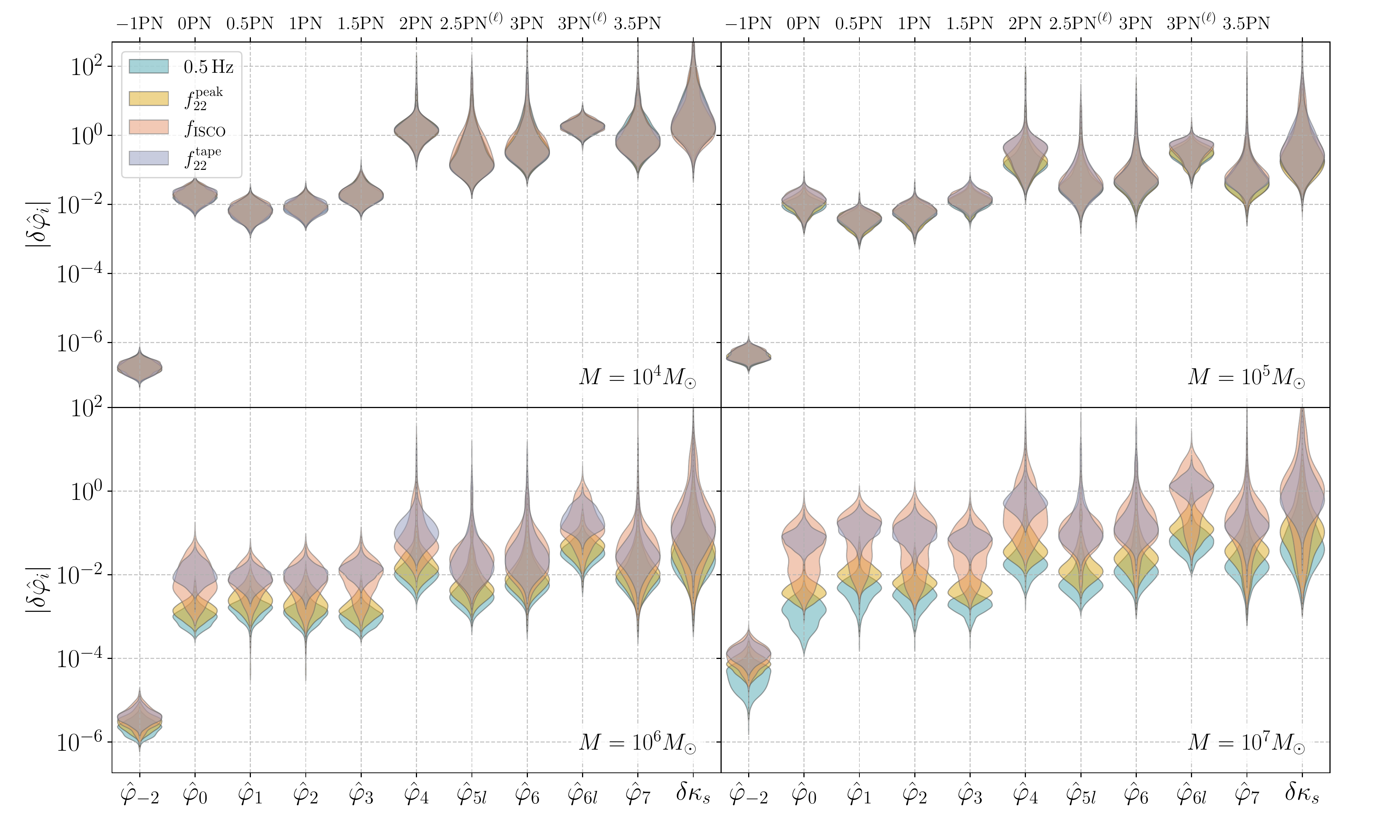}
    \caption{Distribution of $90\%$ upper bounds for each non-GR deviation parameter $|\delta\hat{\varphi}_i|$ for different total mass systems ($M = 10^4, 10^5, 10^6, 10^7 M_\odot$) using different frequencies cutoffs. The analysis includes four different frequency cutoffs: a fixed value of 0.5~Hz (turquoise), the peak frequency of the (2,2)-mode $f_{22}^{\text{peak}}$ (yellow), the innermost stable circular orbit frequency $f_{\text{ISCO}}$ (salmon), and the tapering-frequency of the (2,2)-mode $f_{22}^{\text{tape}} = 0.35f_{22}^{\text{peak}}$ (violet).} \label{fig:violin_diffM_cutf_comparison}
    \end{figure*}

Figure~\ref{fig:violin_diffM_cutf_comparison} presents a comprehensive set of Fisher uncertainty constraints across the different mass regimes we consider, using the same randomization over other parameters as before, and across frequency cutoffs, revealing insights into the information content of these signals.

First, for low-mass systems ($M = 10^4 M_\odot$), we observe near-identical results across different frequency cutoffs.    
This expected uniformity stems from the fact that the merger-ringdown portion occurs at frequencies well beyond LISA's sensitivity band ($f > 0.5$~Hz), as illustrated in Fig.~\ref{fig:mbh_amp_cutf}, making the signals purely inspiral signals.

Intermediate-mass systems ($M = 10^5 M_\odot$) are more informative. 
Despite the merger occurring within LISA's band, the constraints remain primarily driven by the inspiral phase. 
This behavior is driven by the fact that the majority of the SNR is accumulated during the prolonged inspiral phase, while the merger and ringdown contribute less significantly.
The minimal variation between different frequency cutoffs suggests that the additional information from the merger-ringdown phase does not substantially improve our ability to constrain non-GR parameters.

The situation changes notably for higher-mass systems ($M = 10^6$ and $10^7 M_\odot$), where we observe appreciable variations between different frequency cutoffs across all PN orders. 
This mass range presents a regime where both the inspiral and merger-ringdown phases contribute significantly to the parameter estimation, despite the deviation from GR being tapered to zero before reaching the merger.
The complete signal analysis (0.5~Hz cutoff) consistently provides tighter constraints, particularly at $M=10^7 M_\odot$, compared to cuts at lower frequency, and this remains true for the all PN order parameters, with the largest improvements at low positive order $n$.

\begin{figure*}[htbp]
    \hspace*{-1cm}
    \includegraphics[width=\textwidth]{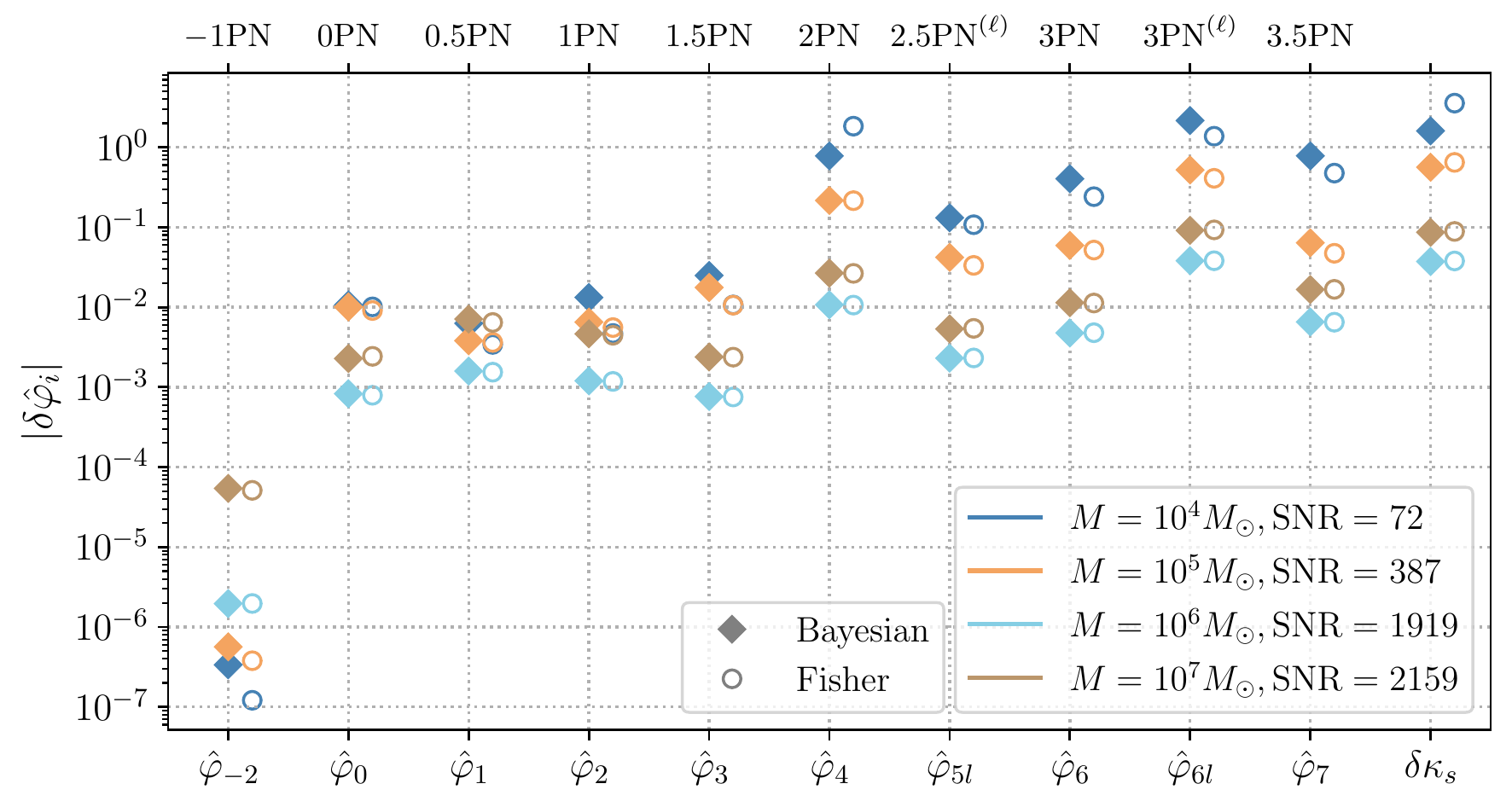}
    \caption{Upper bounds at the $90\%$ level of the deviation parameters $|\delta \hat{\varphi}_i|$ for the four MBHBs signals listed in Table~\ref{tab:param_sources}, computed with both Bayesian (diamonds) and Fisher (circles) methods. The SNR of each system is also listed in the figure.} \label{fig:prod_05_sigma_plot_comparison}
\end{figure*}

The enhanced measurement precision results from complementary information about the system parameters encoded in the merger-ringdown phase, which becomes increasingly valuable as the inspiral phase shifts to lower frequencies where LISA's sensitivity decreases. 
The inclusion of the merger-ringdown part, although modelled in GR, proves valuable for breaking degeneracies between intrinsic and non-GR parameters.
The high-SNR merger-ringdown information leads to a precise estimation of intrinsic parameters, which in turn reduces uncertainties in non-GR parameters through the correlations present in the multidimensional posterior.
The systems providing optimal constraints on certain non-GR parameters require a careful consideration of all phases of the binary coalescence.

In turn, this leads to a \textit{word of caution}. The inspiral-only test that we consider is phenomenological, and the tapering of the effect to impose a GR merger-ringdown is in fact arbitrary. It is possible that true modifed-gravity signals could not be well represented by this construction, in which case our results for high masses show that the impact of this mismodelling might be significant. 
If we were to impose that we only consider the part of the signal where our parametrized deviation has physical meaning, then we should quote measurement errors from the inspiral-only setting of Fig.~\ref{fig:violin_diffM_cutf_comparison} with, for instance, a cut at $f_{22}^{\rm tape}$. On the other hand, cutting away the merger-ringdown and possibly the late inspiral would come at a significant worsening of the constraints for high-mass systems. Assessing further the safety of using an inspiral-only deviation for an inspiral-merg-ringdown analysis would require injecting modified-GR signals with deviations in the merger-ringdown, for instance following the framework of~\cite{Maggio:2022hre, Toubiana:2023cwr}.
    
\subsection{Bayesian Analyses}
\label{sec:results_bayesian}

To validate the Fisher analyses presented in the previous section, we now report the results of our Bayesian parameter estimation on the MBHB systems listed in Table~\ref{tab:param_sources}, using the methodology described in Sec.~\ref{sec:method}. 
For this comparison, the analyses are performed with the \texttt{IMRPhenomXHM} waveform model, and use the IMR setting ($f_{\text{max}} = 0.5\,\text{Hz}$, to capture the full waveform) with tapering off the GR deviation at $f = f^{\text{tape}}$.

The priors adopted in our analysis are as follows: a uniform prior for the chirp mass and the mass ratio, a uniform prior for the luminosity distance, the spins, and also for the angular parameters. 
In particular, the angle pairs $(\iota, \varphi)$ and $(\lambda, \beta)$ are assumed to be uniformly distributed over the sphere, while the polarization angle $\psi$ is assigned a flat prior.
We use uniform priors as well for the non-GR deviation parameters.

\begin{table}[h!]
    \begin{tabularx}{0.497\textwidth}{|c|c|}
        \hline
        Parameter & Injected Value \\
        \hline
        Total Redshifted Mass ($M_z$) [$M_\odot$] & $10^4, 10^5, 10^6, 10^7$ \\
        Mass ratio ($q$) & 3 \\
        Redshift ($z$) & 1\\
        Luminosity Distance ($d_L$) [Mpc] & 6791.8\\
        Dimensionless spin parameters ($\chi_1, \chi_2$) & -0.45, 0.2\\
        Inclination ($\iota$) [rad] & 2.4127\\
        Phase ($\varphi$) [rad] & 0.5251\\
        Ecliptic longitude ($\lambda$) [rad] & -2.5523\\
        Ecliptic latitude ($\beta$) [rad] &  -0.3493\\
        Polarization ($\psi$) [rad] & 1.4014\\
        GR deviation parameter ($\delta\hat{\varphi}_i$) & 0 \\
        \hline
    \end{tabularx}
    \caption{Injected parameter values suitable for Bayesian analysis done in Sec.~\ref{sec:results_bayesian}.}
    \label{tab:param_sources}
\end{table}

\begin{figure*}[htbp]
    \hspace*{0.7cm}
    \includegraphics[width=\textwidth]{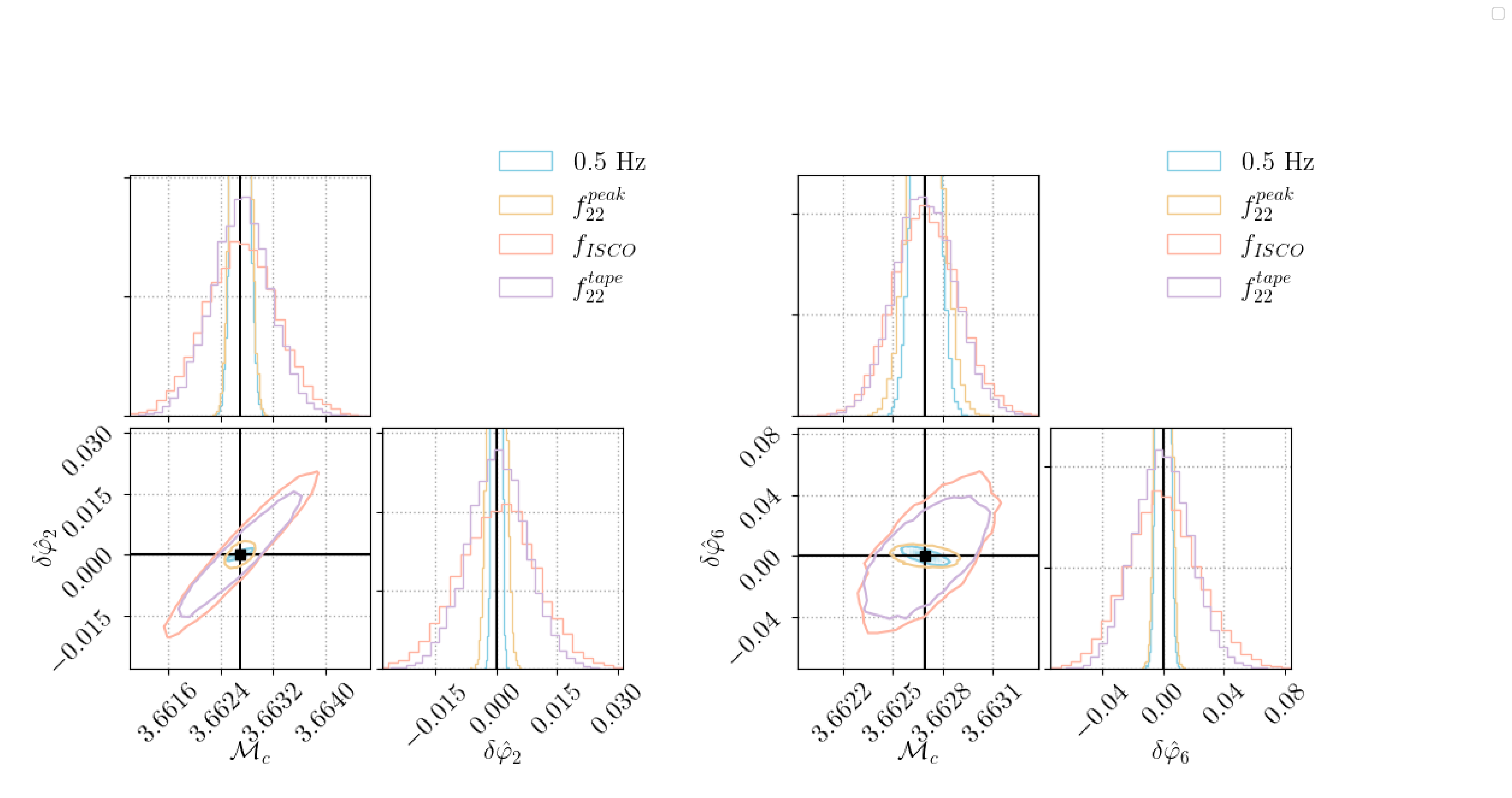}
    \caption{On the left (right) 2D posterior distributions, evaluated with the Bayesian method, of the deviation parameters $\delta \hat{\varphi}_2$ ($\delta \hat{\varphi}_6$) versus $\mathcal{M}_c$, for the MBHBs system listed in Table~\ref{tab:param_sources} with total mass $M = 10^6M_{\odot}$. The analyses are done with the \texttt{IMRPhenomXHM} waveform, using four different frequency cutoffs: a fixed value of 0.5~Hz (turquoise), the peak frequency of the (2,2) mode $f_{22}^{\text{peak}}$ (yellow), the innermost stable circular orbit frequency $f_{\text{ISCO}}$ (salmon), and the tapering-frequency of the (2,2) mode, $f_{22}^{\text{tape}} = 0.35f_{22}^{\text{peak}}$ (violet).} 
    \label{fig:combined_corner_plot_cutf_fti_dchi2v_fti_dchi6v}
    \end{figure*}
    
\begin{figure*}[htbp]
    \hspace*{-1cm}
    \includegraphics[width=\textwidth]{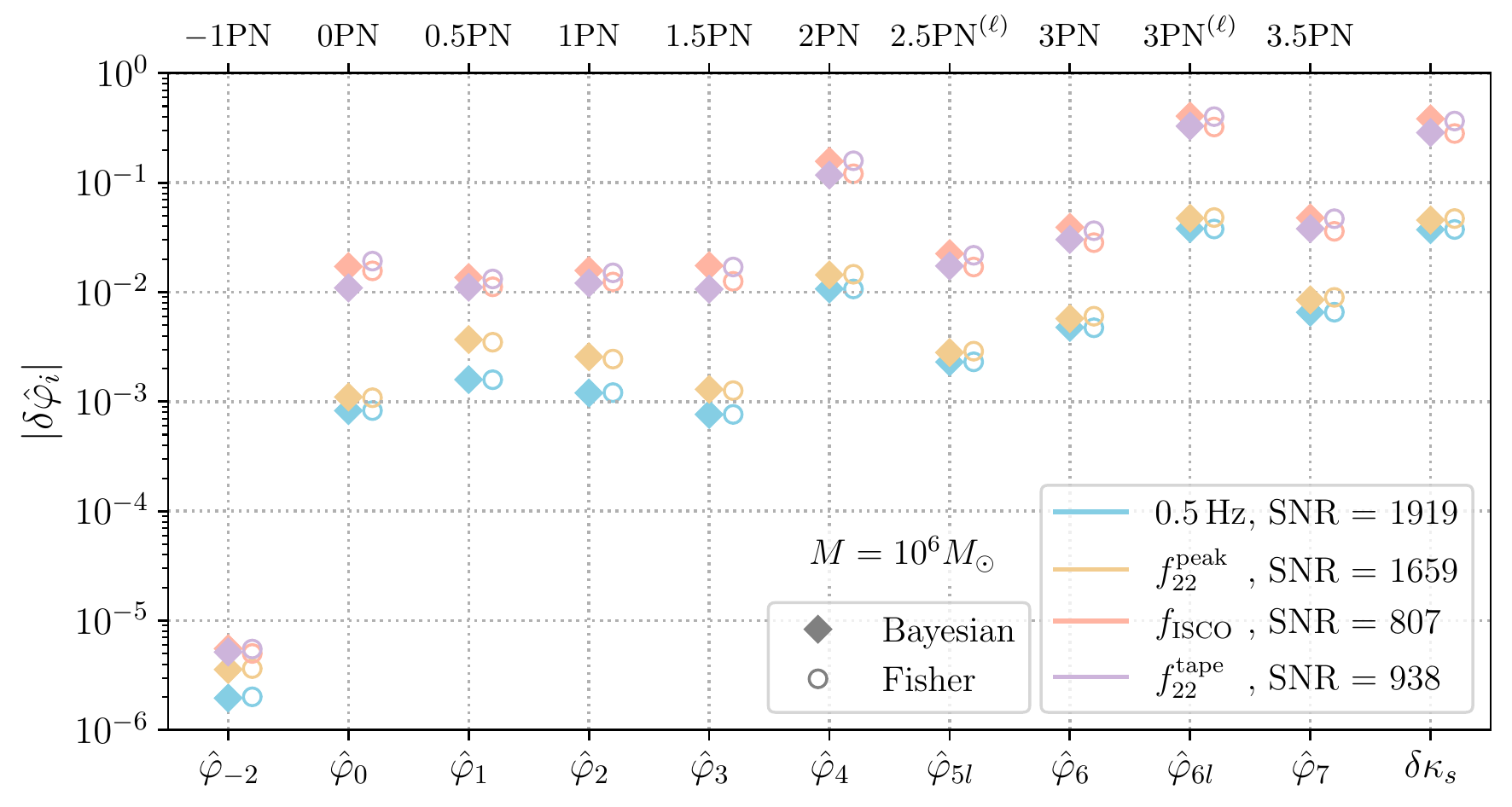}
    \caption{90\% upper bounds on the deviation parameters, $|\delta \hat{\varphi}_i|$, for the MBHB signal with total mass $M = 10^6 M_{\odot}$, and other parameters listed in Table~\ref{tab:param_sources}. The results are shown for different values of maximal frequency $f^\text{{max}}$ at which the data are cut, in accordance to Eq~\eqref{eq:inner_prod_f_max}.  The SNR of the signal per each frequency cut is also listed in the figure.}.
    \label{fig:sigma_plot_cut_f}
    \end{figure*}

Figure~\ref{fig:prod_05_sigma_plot_comparison} presents the 90\% upper bounds for the non-GR parameters $|\delta \hat{\varphi}_i|$ obtained from our Bayesian analysis, and the corresponding Fisher uncertainties.

These results consistently exhibit the same trends across all PN orders as those observed in our previous Fisher-based analyses. 
Although the $90\%$ upper bounds obtained from the Fisher matrix estimates fall quite close to the Bayesian inference result, we find that some caution is warranted for low-mass systems: for sources with total masses around $10^4\,M_{\odot}$, we observe mild discrepancies between the Bayesian and Fisher results across most deviation parameters.
This deviation likely reflects the breakdown of the Gaussian approximation in the low-mass regime, where the assumptions underlying the Fisher matrix approach — such as high SNR and approximately Gaussian likelihoods — are less well satisfied. It reflects also the case that for inspiral-only signals, a stronger hierarchy exists between intrinsic parameters (with the chirp mass $\mathcal{M}_c$ and spin combination $\chi_{\rm PN}$ driving the phasing), with more degeneracies affecting the recovery of subdominant parameters.
    
Conversely, for mass systems with $M \geq 10^5 M_{\odot}$ the higher SNR and inspiral-merger-ringdown signal morphology ensures that Fisher-based estimates remain robust and closely match results from full Bayesian inference. 
Although we did not validate the Fisher matrix over a full population of MBHBs, it seems that for these masses, it serves as an excellent tool for rapid parameter uncertainty estimation, in agreement with more detailed Bayesian analyses (for the circularized, aligned-spin case that we consider).
    
To validate as well the Fisher analysis results presented in Sec.~\ref{sec:cut_f}, where we studied the impact in the analyses of different upper frequency cutoffs, we performed a similar Bayesian inference study on a representative MBHB system of total mass $ M = 10^6 M_{\odot} $, with parameters listed in Table~\ref{tab:param_sources}.

The posterior distributions were found to be consistent with the Fisher matrix predictions, with example posteriors shown for the $\delta \hat{\varphi}_2$ and $\delta \hat{\varphi}_6$ parameters in Fig.~\ref{fig:combined_corner_plot_cutf_fti_dchi2v_fti_dchi6v}.
These examples show how the constraints tighten significantly as more of the merger-ringdown signal is retained — not only for higher PN order parameters like $\delta \hat{\varphi}_6$, which naturally benefit from the high-frequency content of the waveform, but also for lower-order coefficients such as $\delta \hat{\varphi}_2$.
While one might expect low-PN parameters to be primarily constrained by the inspiral, this behavior reflects the role of the merger-ringdown phase in breaking degeneracies with intrinsic binary parameters (such as the chirp mass $\mathcal{M}_c$), leading to tighter constraints on the non-GR coefficients as well.

The comparisons for this system are summarized in Fig.~\ref{fig:sigma_plot_cut_f}, which reports the 90\% credible upper bounds on the absolute value of each $|\delta \hat{\varphi}_i|$ as a function of the maximal frequency cutoff for both Bayesian (diamonds) and Fisher analyses (circles). 
The full-signal analysis (with a 0.5~Hz cutoff) yields the tightest constraints, in line with the discussion in Sec.~\ref{sec:cut_f}, where we noted that this mass regime benefits from contributions of both inspiral and merger-ringdown phases to the parameter estimation. 

The close agreement between the Bayesian posteriors and the Fisher estimates validates the use of the Fisher approach for systems of this type in the case where a maximal frequency cut is used, which is an important validation as those posteriors correspond to a lower SNR, and are more extended with more room for parameter degeneracies.

\section{\label{sec:conclusions}Summary and Conclusions}

In this work, we extended and implemented the FTI framework within the \texttt{lisabeta} package, to investigate LISA's capability to constrain inspiral deviations from GR using MBHB systems across a broad range of total masses ($10^4$ to $10^7 M_\odot$).
The framework introduces correction terms $\delta\hat{\varphi}_i$ to the PN phase coefficients from $n = -2$ to $n = 7$, capturing potential beyond-GR modifications while restricting deviations to the inspiral regime through a smooth tapering to zero before merger. 
We then applied both Fisher information and Bayesian analyses to quantify LISA’s sensitivity to these deviations.

Our results confirm that MBHB observations by LISA will provide unprecedented constraints on non-GR deviation parameters, with improvements of at least two orders of magnitude compared to current single-source constraints from ground-based detectors.
Most striking is the enhancement for the $-1$PN order parameter $\delta\hat{\varphi}_{-2}$, showing improvements of approximately four orders of magnitude for systems in the $[10^4, 10^5] M_\odot$ mass range. 
These substantial improvements arise from the extended duration of MBHB signals in LISA's sensitivity band and the large number of accumulated orbital cycles at low frequencies.
The projected LISA bounds on the spin-induced multipole parameter $\delta\kappa_s$, improving over current ground-based limits by several orders of magnitude for high-mass binaries ($10^6$–$10^7 M_{\odot}$), highlight its capability to probe the nature of compact objects and to discriminate BHs from exotic alternatives.

Our analysis reveals optimal performance for systems with total masses between $10^5$ and $10^6 M_\odot$, representing an ideal compromise between long inspiral phases and high SNR. 

We explored the relative information content of the inspiral and merger-ringdown through an analysis cutting the signal at different maximum frequencies, and found that for higher-mass systems ($M \gtrsim 10^6 M_\odot$), the inclusion of merger-ringdown information significantly enhances parameter constraints by breaking degeneracies between intrinsic and non-GR parameters.
This finding highlights that incorporating the late-inspiral and merger-ringdown plays an important role in improving the precision of the constraints; on the other hand, it also shows that the current phenomenological setup of this inspiral-only test—with an artificial tapering before merger—may be problematic if the assumption of a GR-like merger–ringdown is inaccurate. Since our analysis set-up does not allow yet for the injection an IMR or merger-ringdown-only deviation, we leave the assessment of this type of systematics for future work.

We validated our Fisher matrix approach through detailed Bayesian analyses, confirming robust performance for systems with $M \geq 10^5 M_\odot$, while identifying the need for full Bayesian inference in low-SNR inspiral-dominated regimes ($M \sim 10^4 M_\odot$).

Overall, our findings establish LISA as a unique instrument for fundamental physics, capable of placing stringent constraints on deviations from GR through MBHB observations.
Our parameterized test framework provides a robust, model-agnostic approach for probing gravitational theory in the strong-field regime.

Our study leaves open a number of questions warranting further investigation.
A comprehensive study of waveform systematics and their impact on constraints of non-GR parameters will be essential for quantifying the precision achievable with our current waveforms. 
Such an analysis would help assessing potential biases and the risk of falsely inferring deviations from GR due to waveform mismodelling, as well as the corresponding accuracy requirements on waveform models.
The extension of our parametrized test framework to include more sophisticated waveforms, including precession, eccentricity and environmental effects, such as line-of-sight acceleration due to third-body interactions, will be important in getting closer to realistic LISA signals.
Extending the present analysis to hierarchical population studies based on realistic MBHB formation and evolution scenarios will provide important insights into the expected population-level constraints on alternative theories of gravity.
Finally, the incorporation of realistic noise realizations and potential noise mismodeling systematics will be crucial for assessing the practical limitations of parameter estimation in operational scenarios including the full complexity of the LISA global fit.

\begin{acknowledgments}

We thank Vasco Gennari, Alberto Mangiagli, Maxence Corman, Arnab Dhani, Jonathan Gair and Elisa Maggio for useful insights and fruitful discussions on this work.
M.Piarulli, S.Marsat and N.Tamanini acknowledge support from the French space agency CNES in the framework of LISA.
M.Piarulli and S.Marsat also acknowledge support from the Procope Partenariat Hubert Curien (PHC) Program.
M.Piarulli further acknowledges support from the Erasmus+ program, the University of Toulouse, and the Max Planck Institute for Gravitational Physics (Albert Einstein Institute). A.Buonanno resarch is  supported in part by the European Research Council (ERC) Horizon Synergy Grant “Making Sense of the Unexpected in the Gravitational-Wave Sky” grant agreement no. GWSky–101167314.

\textit{Software}:
We acknowledge usage of the following additional
\textsc{Python} 
packages for modeling, analysis, post-processing, and production of results throughout:
\textsc{matplotlib}~\cite{2007CSE.....9...90H},
\textsc{numpy}~\cite{2020Natur.585..357H},
\textsc{scipy}~\cite{2020NatMe..17..261V}
\textsc{seaborn}~\cite{seaborn}.
\end{acknowledgments}

\appendix

\section{PN phasing and stationary phase approximation}
\label{app:appendix_spa}

In this Appendix, we will give more details on the SPA, and on the structure of the phasing formula ~\eqref{eq:psi_GR} in the PN regime. 
We will use the following Fourier sign convention:
\begin{equation}
    \tilde{h}(f) = \int e^{-2 i \pi f t} h(t) \, dt \,.
\end{equation}

We first consider a generic signal $h(t) = a(t) \exp[i \phi (t)]$ with an increasing phase $\dot{\phi} = \omega > 0$.
Within the SPA, the Fourier-domain signal will have support on positive frequencies $f > 0$ for $\dot{\phi} > 0$, and on negative frequencies for $\dot{\phi} < 0$.
In the standard conventions for spin-weighted spherical harmonics, $h_{\ell m} \propto \exp[- i m \phi_{\rm orb}]$ with $\phi_{\rm orb}$ the orbital phase ($\omega_{\rm orb} = \dot{\phi}_{\rm orb} > 0$).
This means that, for instance, the quadrupolar mode $h_{2,-2}$ will be the one to have support on $f>0$, not $h_{22}$.
This technical point should be remembered as the literature often refers to the dominant harmonic as the (2,2)-mode.

Under the assumptions $|\dot{a}/(a\omega)| \ll 1$, $|\ddot{\omega}/\omega^3| \ll 1$ and $| \dot{a}^2/(a^2 \dot{\omega}) | \ll 1$ that are well verified in the PN regime, the SPA gives $\tilde{h}(f) = A(f) \exp[i \psi(f)]$ with
\begin{align}
    A(f) &= a(t_f) \sqrt{\frac{2\pi}{\dot{\omega}(t_f)}} \,, \nonumber \\
    \psi(f) &= \phi(t_f) - 2\pi f t_f + \frac{\pi}{4} \,,
\end{align}
where the time $t_f$ identifies the time-to-frequency correspondence along the orbital evolution, i.e. $t_f$ is such that $\omega(t_f) = 2 \pi f $. Differentiating the phase $\psi(f)$, we also obtain
\begin{equation}
    t_f = -\frac{1}{2\pi} \frac{d\psi}{df} \,,
\end{equation}
while differentiating again yields the for radiation-reaction timescale $T_f$ (see Ref.~\cite{SPA_Sylvain})
\begin{equation}\label{eq:def_Tf}
    T_f^2 = \frac{1}{\dot{\omega}(t_f)} = -\frac{1}{4\pi^2} \frac{d^2\psi}{df^2} \,.
\end{equation}

We now specialize the above to a waveform with multiple harmonics $h_{\ell m}$.
We assume a spin-aligned (and circularized) system with no spin precession, for which the symmetry relation $h_{\ell, -m} = (-1)^{\ell} h_{\ell m}^{*}$ allows us to consider only $m > 0$ modes, having support on $f > 0$.
In the PN regime, the different modes with amplitudes $a_{\ell m}$ and phases $\phi_{\ell m}$ will follow the structure
\begin{equation}\label{eq:philm_structure}
    \phi_{\ell m} = -m \phi_{\rm orb} + \Delta \phi_{\ell m} \,,
\end{equation}
with $\phi_{\rm orb}$ the orbital phase and where the $\Delta \phi_{\ell m}$ are slowly varying with orbital frequency, and take values $\in \mathbb{Z} \pi/2$ in the limit $\omega_{\rm orb} \rightarrow 0$, which can be read off the leading PN order of mode amplitudes (see Ref.~\cite{2024LRR....27....4B}); we will treat them as constant in what follows.
Applying the SPA to the different modes, the time-to-frequency correspondence differs for each, with $-m \omega_{\rm orb}(t_f^{\ell m}) = 2\pi f$. However, because of the structure~\eqref{eq:philm_structure}, one can derive the scaling relations $t^{\ell m}(-m f/2) = t^{2,-2}(f)$ and
\begin{align}\label{eq:Psilm_scaling}
    2\psi_{\ell m} \left(\frac{-m f}{2}\right) =& -m \psi_{2,-2}(f) + (m+2)\frac{\pi}{4} \nonumber\\
    &+ m\Delta \phi_{2,-2} + 2\Delta \phi_{\ell m} \,.
\end{align}

In the PN regime, for circularized binaries the inspiral phasing is obtained from the balance equation $\dot{E} = -\mathcal{F}$ between the orbital energy $E(v)$ and the energy flux emitted in gravitational waves $\mathcal{F}(v)$, that we will both write as functions of the PN parameter $v = (M \omega_{orb})^{1/3}$ ($G=c=1$), with a $v^2$ factor corresponding to 1PN corrections. Defining
\begin{equation}\label{eq:def_Gofv}
    \mathcal{G}(v) = -\frac{E'(v)}{\mathcal{F}(v)} = \frac{dt}{dv} \,,
\end{equation}
where the $'$ means a derivative with respect to $v$, we have
\begin{align}\label{eq:PN_phasing}
    t(v) &= \int^{v} dv \, \mathcal{G}(v) \,, \nonumber \\
    \phi_{\rm orb}(v) &= \frac{1}{M} \int^{v} dv \, v^3 \mathcal{G}(v) \,.
\end{align}
This defines $t(v)$ and $\phi_{\rm orb}(v)$ up to integration constants, which are fixed by imposing values $t_0$, $\phi_0$ as some reference orbital frequency $v_0$. We are following here the TaylorF2 prescription (see Ref.~\cite{2009PhRvD..80h4043B} for a nomenclature), where the right-hand-side of Eq.~\eqref{eq:PN_phasing} is re-expanded at a given PN order as a perturbative series in $v$ before being integrated term-by-term. The result can be translated into the phasing for modes $h_{\ell m}$ using the SPA, with the correspondence $2\pi f = -m v^3/M$. For the leading quadrupolar mode, this gives in GR
\begin{equation}
    \psi_{2,-2}(f) = \frac{3}{128 \eta} v^{-5} \left[ 1 + (\dots)v^2 + \dots \right] + \mathrm{const} \,,
\end{equation}
where $v = (\pi M f)^{1/3}$. Explicitly, if we use the notation $T(v)$ and $\Phi(v)$ for the formal, term-by-term integration of the $v$-series in the right-hand-side of Eq.~\eqref{eq:PN_phasing} while setting the integration constant to zero, and if we enforce $t(v_0) = t_0$ and $\phi_{\rm orb}(v_0) = \phi_0$ at some reference $v_0$, then we have
\begin{align}
    t(v) - t_0 &= T(v) - T(v_0) \,, \nonumber \\
    \phi_{\rm orb}(v) - \phi_0 &= \Phi(v) - \Phi(v_0) \,,
\end{align}
and
\begin{align}\label{eq:psi22_align_explicit}
    \psi_{2,-2}(f) =& 2 \Phi(v) - \frac{2v^3}{M} T(v) \nonumber\\
    &- \frac{2v^3}{M} \left(t_0 - T(v_0) \right) + 2 \left(\phi_0 - \Phi(v_0) \right) \nonumber\\
    &+ \frac{\pi}{4} + \Delta \phi_{2,-2} \,.
\end{align}
The first two terms correspond to~\eqref{eq:psi_GR} and are often the only ones written explicitly, as waveform models enforce alignment using different conventions.

Here we ignored a constant term (alignment phase $\phi_0$, constant $\pi/4$, possible term $\Delta \phi_{2,-2}$), as well as a linear term corresponding to the time at alignment $t_0$. Note that the constant and linear terms are respectively degenerate with the $v^5$ and $v^8$ corrections in the PN series.

These constants and alignment conventions are handled by the underlying waveform model representing the GR waveform, which also takes care of the relative mode phases appearing in~\eqref{eq:Psilm_scaling}. We are treating the extra phase contribution coming from the deviation from GR as an additive dephasing $\delta \psi_{\ell m}(f)$. 
Consistent with~\cite{FTI}, we taper the deviation from GR before reaching the merger part of the signal, by applying a window directly to the second derivative of the Fourier-domain phase modification, as written in~\eqref{eq:phase_correction}, which we reproduce here:
\begin{equation}
    \delta\psi_{2,-2}(f) = \int_{f_{\text{int}}}^f \! df' \int_{f_{\text{int}}}^{f'} \! df'' \, 
    \delta\psi''_{2,-2}(f'') W(f'').
\end{equation}
Here, $\delta\psi''_{2,-2}$ is the second derivative with respect to frequency. The correspondence~\eqref{eq:def_Tf} justifies this choice: since $d^2\psi/df^2 \propto 1/\dot{\omega}_{\rm orb} \propto$ and $\mathcal{G}(v) = dt / dv = 3v^2 / (M \dot{\omega}_{\rm orb})$ according to~\eqref{eq:def_Gofv}, we see that we are applying directly the tapering on the deviation to the quantity $\mathcal{G}$\footnote{This is up to a change of variables from $v$ to $f$; in practice, the code implements a Planck window as a function of $f$; the functional shape would be slightly different when seen as a function of $v$.}, as
\begin{equation}
    \mathcal{G} + \Delta \mathcal{G} \rightarrow \mathcal{G} + W\Delta \mathcal{G} \,.
\end{equation}
Since $\mathcal{G}(v) = -E'(v)/\mathcal{F}(v)$, this allows us to apply the deviation from GR, and its tapering to zero, directly on physically meaningful quantities, free of any alignment ambiguity. 
The conventional choice of alignment, deciding a reference frequency at which the time and phase difference induced in the final waveform are set to vanish, is decided by the integration bounds in~\eqref{eq:phase_correction}. 
This procedure requires integrating numerically twice, which can be made computationally efficient as explained in the main text.

Finally, the phase correction $\delta\psi_{2,-2}$ is rescaled to be applied to the other modes. 
We can apply the correction $\Delta \mathcal{G}$ in~\eqref{eq:PN_phasing} and follow the same steps leading to the scaling~\eqref{eq:Psilm_scaling} for only the corrections $\delta \psi$, which gives
\begin{equation}
    \delta \psi_{\ell m}\left( \frac{-m f}{2} \right) = \frac{-m}{2}\delta \psi_{2,-2}(f) \,.
\end{equation}
With $\delta \psi_{\ell m}$ given as a power series in the orbital $v$, it is useful to recall the relation to orbital frequency through the relation $v = (-2\pi M f/m)^{1/3}$. 
This rescaling means that the numerical integration in~\eqref{eq:phase_correction} needs only to be computed once, with the rescaling propagating the change to other modes.

\section{Role of the tapering parameter}
\label{app:appendix_alpha}

In this Section, we study the impact of different values of $\alpha$ for the tapering frequency, $f_{\text{tape}}^{22} = \alpha f_{\text{peak}}^{22}$, on our analysis.
The impact of different tapering frequencies reveals additional insights into the robustness of our parameter estimation across the parameter space, and the level of arbitrariness induced by the choice of $\alpha$.
By varying $\alpha = 0.35, 0.5, 0.7, 0.9$, we explore how the transition between non-GR and GR waveform affects our ability to constrain non-GR parameters.
This study closely follows the methodology used in our frequency cutoff analysis in Sec.~\ref{sec:results_fisher}.

\begin{figure*}[ht!]
    \centering
    \includegraphics[width=\textwidth]{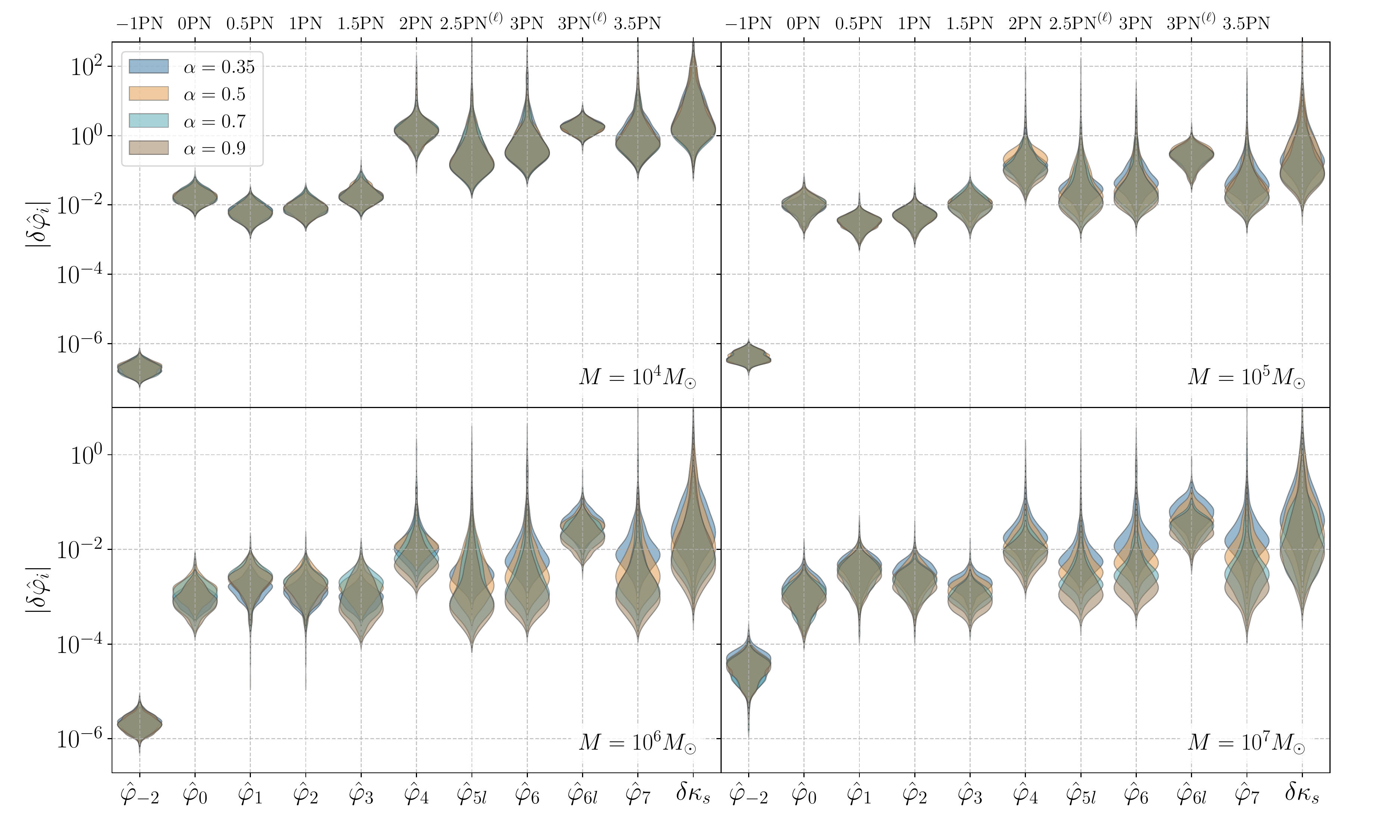}
    \caption{Distribution of $90\%$ upper bounds for each non-GR deviation parameter $|\delta\hat{\varphi}_i|$ for different total mass systems ($M = 10^4, 10^5, 10^6, 10^7 M_\odot$) using different values of the tapering frequency, where the GR deviation is tapered off. For each modes $f_{lm}^{\text{tape}} = m/2 \cdot f^{\text{tape}}_{22} = m/2 \cdot \alpha f^{\text{peak}}_{22} $.} \label{fig:violin_diffM_alpha_comparison_SEOBv5}
    \end{figure*}

We start by showing our results computed with Fisher matrix analyses, summarized in Fig.~\ref{fig:violin_diffM_alpha_comparison_SEOBv5}.
For low-mass systems ($M = 10^4 M_\odot$) and intermediate mass systems ($M = 10^5 M_\odot$), the constraints show minimal variation across different $\alpha$ values. 
The independence from $\alpha$ values reinforces our understanding that the parameter estimation for these systems is primarily driven by the high number of cycles in band, during the long inspiral phase.
The most pronounced effects of varying $\alpha$ appear in the high-mass systems ($M = 10^6$ and $10^7 M_\odot$). 
Here, we observe clear differences in the constraint patterns across all PN orders. 
For these systems, where the merger phase occurs in LISA's most sensitive band, the choice of tapering frequency directly impacts how much of the late-inspiral and early-merger information is included in the analysis. 
Notably, higher $\alpha$ values (0.7-0.9) tend to produce tighter constraints, particularly for higher PN order parameters, suggesting that including more of the late-inspiral phase enhances our ability to test GR deviations.

This mass-dependent sensitivity to $\alpha$ values aligns with our previous findings regarding frequency cutoffs, reinforcing the importance of carefully considering the transition between inspiral and merger phases in parameter estimation studies, particularly for high-mass systems where this transition occurs within LISA's optimal sensitivity band.

Interestingly, for systems with a total mass of $M = 10^6 M_{\odot}$, we observe a counterintuitive trend at low-PN orders (0, 0.5, 1, 1.5). 
As $\alpha$ increases from 0.35 to 0.9, the $90\%$ credible intervals, $|\delta\hat{\varphi}_i|$, do not systematically improve, despite the inclusion of additional data in the non-GR regime. 
This behavior is likely due to strong correlations between these parameters and the intrinsic binary properties, with the introduction of an earlier tapering leading to degeneracy breaking.

\begin{figure*}[ht!]
    \centering
    \includegraphics[width=0.8\textwidth]{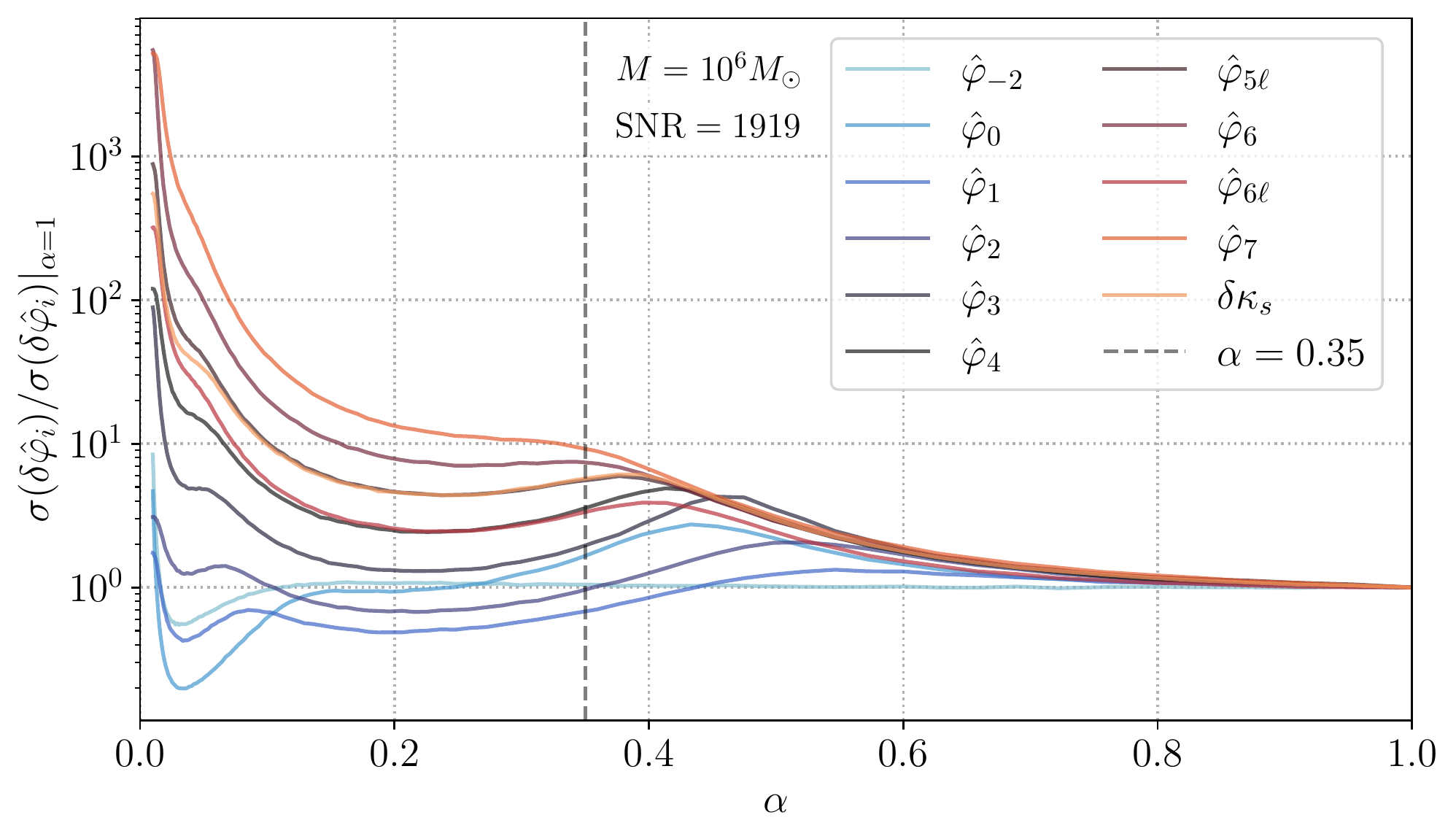}
    \caption{Relative constraint precision $\sigma(\delta\hat{\varphi}_i)/\sigma(\delta\hat{\varphi}_i)|_{\alpha=1}$ as a function of the tapering parameter $\alpha$ for a MBHB system with total mass $M = 10^6 M_{\odot}$, the others parameters are listed in Table~\ref{tab:param_sources}. The analyses are based on Fisher estimates. Each curve represents a different non-GR deviation parameter $\delta\hat{\varphi}_i$, showing how the constraint strength varies with the choice of tapering frequency. The constraints are normalized by their values at $\alpha = 1$ to facilitate comparison across different parameters. The vertical dashed line indicates our fiducial choice of $\alpha = 0.35$.}
    \label{fig:sigma_alpha_M6}
\end{figure*}

Figure~\ref{fig:sigma_alpha_M6} provides a detailed view of the Fisher uncertainties for individual non-GR parameters as a function of the tapering parameter $\alpha \in [0.01, 1]$, for the representative $M = 10^6 M_{\odot}$ system, listed in Table~\ref{tab:param_sources}. 
The figure reveals a complex, parameter-dependent relationship between the tapering frequency and constraint strength that is not always monotonic with $\alpha$. 
As expected, small $\alpha$ values result in constraints that become extremely poor, indicating that it becomes impossible to place meaningful constraints on non-GR parameters. 
Since $f_{\text{tape}}^{22} = \alpha f_{\text{peak}}^{22}$ determines the frequency to which non-GR deviations are tapered off, very small $\alpha$ values limit the frequency range where deviations from GR are allowed, making parameter estimation ineffective. 
Not surprisingly most of the non-GR parameters show improvements at higher $\alpha$ values, consistent with our earlier findings that they benefit from including more data in the analyses.
This non-monotonic behavior emphasizes the complex interplay between information gain and parameter degeneracies as more signal content is included. 

To validate our Fisher analysis findings, we also performed a Bayesian inference analysis on the same representative MBHB system with total mass $M = 10^6 M_{\odot}$, examining the impact of different $\alpha$ values ($\alpha = 0.35, 0.5, 0.7, 0.9$) on the 90\% upper bounds of the deviation parameters $|\delta \hat{\varphi}_i|$. 
The results showed a good consistency between the Fisher and Bayesian results, across all values of alpha, with a level of agreement similar to that shown in Fig.~\ref{fig:sigma_plot_cut_f}.

This analysis reinforces that our choice of $\alpha = 0.35$ provides somewhat conservative constraints, but also underlines the level of arbitrariness present in this IMR test for inspiral deviations.
This value $\alpha = 0.35$ ensures that our constraints on non-GR parameters are robust and less susceptible to potential systematic errors that might arise from imperfect modeling of the late-inspiral phase; however, the improved constraints observed with higher $\alpha$ values suggest that there may be untapped potential for enhanced parameter estimation.

\newpage
\vfill

\bibliographystyle{apsrev4-2} 
\bibliography{main}

\end{document}